\documentclass[
prd,
aps,
superscriptaddress,
groupedaddress,
showpacs,
preprintnumbers,
amsmath,
floatfix,
twocolumn,
nofootinbib
]{revtex4-1}

\usepackage{epsfig}
\usepackage{latexsym}
\usepackage[psamsfonts]{amssymb}
\usepackage{graphicx}
\usepackage{epstopdf}
\usepackage{bm}
\usepackage{color}
\usepackage{CJK}

\begin{document}
\bibliographystyle{apsrev4-1}
\preprint{Edinburgh 2010/09, KEK-TH-1338,  NT@UW-10-04, RBRC-821, TKYNT-10-03, UTCCS-P-58}

\begin{CJK*}{UTF8}{}

\title{Nucleon isovector structure functions in (2+1)-flavor QCD with domain wall fermions}

\CJKfamily{min}
\author{Yasumichi Aoki (青木保道)} 
\affiliation{RIKEN-BNL Research Center, Brookhaven National Laboratory, Upton, NY 11973}

\author{Tom Blum} 
\affiliation{Physics Department, University of Connecticut, Storrs, CT 06269-3046}

\CJKfamily{bsmi}
\author{Huey-Wen Lin (林慧雯)}
\affiliation{Department of Physics, University of Washington, Seattle, WA 98195-1560}

\CJKfamily{min}
\author{Shigemi Ohta (太田滋生)
} 
\affiliation{Institute of Particle and Nuclear Studies, KEK, Tsukuba, Ibaraki 305-0801, Japan}
\affiliation{Physics Department, Sokendai Graduate University of Advanced Studies, Hayama, Kanagawa 240-0193, Japan}
\affiliation{\,RIKEN-BNL Research Center, Brookhaven National Laboratory, Upton, NY 11973}

\author{Shoichi Sasaki (佐々木勝一)}
\affiliation{Department of Physics, University of Tokyo, Hongo 7-3-1, Tokyo 113-0033, Japan}

\author{Robert Tweedie and James Zanotti}
\affiliation{School of Physics \& Astronomy, The University of Edinburgh, Edinburgh EH9 3JZ, UK}

\author{Takeshi Yamazaki (山崎剛)}
\affiliation{Center for Computational Sciences, University of Tsukuba, Tsukuba, Ibaraki 305-8577, Japan}


\collaboration{RBC and UKQCD Collaborations}


\date{
March 17, 2010
}

\begin{abstract}
We report on numerical lattice QCD calculations of some of the low moments of the nucleon structure functions.
The calculations are carried out with gauge configurations generated by the RBC and UKQCD collaborations with (2+1)-flavors of dynamical domain wall fermions and the Iwasaki gauge action ($\beta = 2.13$). The inverse lattice spacing is $a^{-1} = 1.73$ GeV, and two spatial volumes of \((2.7\;{\rm fm})^3\) and \((1.8\;{\rm fm})^3\) are used. 
The up and down quark masses are varied so the pion mass lies between 0.33 and 0.67 GeV while 
the strange mass is about 12 \% heavier than the physical one.
The structure function moments we present include fully non-perturbatively renormalized  iso-vector quark momentum fraction, \(\langle x \rangle_{u-d}\), helicity fraction, \(\langle x \rangle_{\Delta u - \Delta d}\), and transversity, \(\langle 1 \rangle_{\delta u - \delta d}\),  as well as an unrenormalized twist-3 coefficient, \(d_1\).
The ratio of the momentum to helicity fractions, \(\langle x \rangle_{u-d}/\langle x \rangle_{\Delta u - \Delta d}\), does not show dependence on the light quark mass and agrees well with the value obtained from experiment.
Their respective absolute values, fully renormalized, show interesting trends toward their respective experimental values at the lightest quark mass.
A prediction for the transversity, \(0.7 < \langle 1 \rangle_{\delta u -\delta d} < 1.1\), in the \(\overline{\rm MS}\) scheme at 2 GeV is obtained.
The twist-3 coefficient, \(d_1\), though yet to be renormalized, supports the perturbative Wandzura-Wilczek relation.
\color{black}
\end{abstract}

\maketitle
\end{CJK*}

\section{Introduction}

We report numerical lattice quantum chromodynamics (QCD) calculations of some low moments of nucleon structure functions using the lattice gauge ensembles \cite{Allton:2008pn} jointly generated by the RIKEN-BNL-Columbia (RBC) and UKQCD Collaborations with ``2+1'' flavors of dynamical domain-wall fermions (DWF)~\cite{Kaplan:1992bt,Shamir:1993zy,Furman:1994ky}.
Recently, there has been an increased interest in lattice calculations of  these moments (see \cite{Zanotti:2008zm,Hagler:2009ni,Renner:2010ks} for recent reviews).

The structure functions are measured in deep inelastic scattering of electrons off a nucleon~\cite{Breidenbach:1969kd,Friedman:1991nq,Kendall:1991np,Taylor:1991ew,Gluck:1995yr,Gehrmann:1995ag,Lai:1996mg,Adams:1997tq,Adeva:1997qz,Gluck:1998xa,Ackerstaff:1999ey,Martin:2001es}, the cross section of which is factorized in terms of leptonic and hadronic tensors,
\(
\propto
l_{\alpha\beta}W^{\alpha\beta}.
\)
Since the electron leptonic tensor, \(l_{\alpha\beta}\), is known, the cross section provides us with structure information about the target nucleon through the hadronic tensor,
\begin{equation}
W^{\alpha\beta} = i \int d^4x e^{iqx} \langle N | T[J^\alpha(x)J^\beta(0)] | N\rangle.
\end{equation}
Here \(q\) denotes the spacelike four-momentum transferred to the nucleon from the electron through a virtual photon.
The hadronic tensor is decomposed into symmetric unpolarized and antisymmetric polarized parts, \( W^{\alpha\beta} = W^{\{\alpha\beta\}}+W^{[\alpha\beta]}\):
\begin{widetext}
\begin{equation}\label{eq:Wsym}
W^{\{\alpha\beta\}}(x,Q^2) =
\left( -g^{\alpha\beta} + \frac{q^\alpha
q^\beta}{q^2}\right)  {F_1(x,Q^2)} +
\left(P^\alpha-\frac{\nu}{q^2}q^\alpha\right)\left(P^\beta-\frac{\nu}{q^2}q^\beta\right)
\frac{F_2(x,Q^2)}{\nu},
\end{equation}
\begin{equation}\label{eq:Want}
W^{[\alpha\beta]}(x,Q^2) = i\epsilon^{\alpha\beta\gamma\delta} q_\gamma \left\{
\frac{S_\delta}{\nu}[{g_1(x,Q^2)} \right. +\left. {g_2(x,Q^2)}] - 
\frac{(q\cdot S) P_\delta}{\nu^2}{g_2(x,Q^2)} \right\},
\end{equation}
\end{widetext}
with kinematic variables defined as \(P_\alpha\) the nucleon momentum, \(S_\alpha\) the spin normalized with the nucleon mass \(M\), \(S^2 = -M^2\), \(\nu = q\cdot P\), \(x=Q^2/2\nu\), and \(Q^2=|q^2|\).
The unpolarized structure functions are \(F_1(x,Q^2)\) and \(F_2(x,Q^2)\), and the polarized, \(g_1(x,Q^2)\) and \(g_2(x,Q^2)\).
Their moments are described in terms of Wilson's operator product expansion:
\begin{widetext}
\begin{eqnarray}\label{eq:moments}
2 \int_0^1 dx\,x^{n} {F_1(x,Q^2)} &=& \sum_{q=u,d}
c^{(q)}_{1,n}(\mu^2/Q^2,g(\mu))\: \langle x^n \rangle_{q}(\mu)
+{O(1/Q^2)},\\
\int_0^1 dx\,x^{n-1} {F_2(x,Q^2)} &=& \sum_{f=u,d}
c^{(q)}_{2,n}(\mu^2/Q^2,g(\mu))\: \langle x^n \rangle_{q}(\mu)
+{O(1/Q^2)},\\
2\int_0^1 dx\,x^n {g_1(x,Q^2)}
  &=& \sum_{q=u,d} e^{(q)}_{1,n}(\mu^2/Q^2,g(\mu))\: \langle x^n \rangle_{\Delta q}(\mu)
+{O(1/Q^2)},\\
\nonumber\\
2\int_0^1 dx\,x^n {g_2(x,Q^2)}
  &=& \frac{1}{2}\frac{n}{n+1} \sum_{q=u,d} \left[e^{q}_{2,n}(\mu^2/Q^2,g(\mu))\: d_n^{q}(\mu)
-  2 e^{q}_{1,n}\: \langle x^n \rangle_{\Delta
q}(\mu)\right] + {O(1/Q^2)}, 
\end{eqnarray}
\end{widetext}
where the Wilson coefficients, \(c_1\), \(c_2\), \(e_1\), and \(e_2\), are perturbatively known.
The moments, \({\langle x^n \rangle_{q}(\mu)}\), \({\langle x^n \rangle_{\Delta q}(\mu)}\) and \(d_n^q(\mu)\) are calculable on the lattice as forward nucleon matrix elements of certain local and gauge-invariant operators.

In addition,  the tensor charge,
\begin{equation}
\langle 1\rangle_{\delta q}(\mu) = 
\frac{M}{2(S_\alpha P_\beta-S_\beta P_\alpha)}
\langle P,S| \overline{q} i \sigma_{\alpha\beta}\gamma_5 q | P, S\rangle,
\end{equation}
which probes the transverse spin structure of the nucleon, will soon be reported by experiments \cite{Anselmino:2007fs,RHICSpin}.
This quantity is calculated on the lattice in much the same way as the DIS structure function moments are calculated.

In this paper we report our lattice numerical calculations of the following four moments of the structure functions: the quark momentum fraction, \(\langle x\rangle_q(\mu)\), the helicity fraction, \(\langle x\rangle_{\Delta q}(\mu)\), the tensor charge, \(\langle 1\rangle_{\delta q}(\mu)\), and twist-3 coefficient \(d_1^q\) of the \(g_2\) polarized structure function.
These are the moments that can be calculated without finite momentum transfer.
For the former three moments, the momentum and helicity fractions and tensor charge, we restrict ourselves to the isovector flavor combination, \(q=u-d\), as this simplifies non-perturbative renormalization.
All three are non-perturbatively renormalized and readily comparable with the corresponding experiments.

The numerical calculations of these moments use the lattice gauge ensembles generated by the RBC and UKQCD Collaborations with ``2+1'' flavors of dynamical domain-wall fermions (DWF).
The good chiral and flavor symmetries of DWF make our calculations and analyses straightforward: in contrast to more conventional fermion formalisms such as staggered or Wilson, there is no question in defining nucleon quantum numbers, nor complications arising from explicit breaking of chiral symmetry.
This advantage is especially important in non-perturbatively renormalizing the results so they can be compared with experiment and phenomenology.
In this paper we report results from the ensembles with lattice cutoff \(a^{-1}\) = 1.73(3) GeV~\cite{Allton:2008pn}.
We consider two spatial volumes, with linear size about 2.7 and 1.8 fm each.
The strange quark mass is fixed at a value about 12\% heavier than its physical value, and the degenerate up and down quark mass is varied for four values from about three quarter to one fifth of the strange quark mass.
Since we only vary the light quark mass in our simulation while the strange quark mass is held fixed, in the following we call the light up and down quark mass \(m_f\), in lattice units, unless explicitly stated otherwise.

The rest of this paper is organized as follows: We explain our computational method in Sec.~\ref{sec:Formulation}.
In Sec.~\ref{sec:Ensembles} we first summarize the numerical lattice QCD ensembles used for this work.
Then we discuss in detail the known systematic errors in the relevant form factors calculated on these ensembles.
The numerical results are presented in section \ref{sec:fractions}.
Finally, we give the conclusions in Sec.~\ref{sec:conclusions}.

We note that some preliminary results from this study were presented in Refs.~\cite{{Yamazaki:2007mk,Ohta:2008kd,Ohta:2009uy}}.

\section{Formulation}
\label{sec:Formulation}

We refer the reader to our recent publications \cite{Orginos:2005uy,Lin:2008uz,Yamazaki:2009zq} and references cited there in for details of our computational method.
Here we give a brief summary for readers' convenience.
We use the standard proton operator, \(B = \epsilon_{abc} (u_a^T C \gamma_5d_b)u_c\) to create and annihilate proton states.
We Gaussian-smear this operator for better overlap with the ground state with both zero and finite momentum.
A Gaussian radius of 7 lattice units was chosen after a series of pilot calculations.
Since the up and down quark mass are degenerate in these calculations, isospin symmetry is exact.
This is of course a well-known good approximation.
We project onto the positive-parity ground state, so our proton two-point correlation function takes the form
\begin{equation}
C_{\rm 2pt}(t) =
\sum_{\alpha,\beta}
\left(\frac{1+\gamma_t}{2}\right)_{\alpha\beta} 
\langle B_\beta(t_{\rm sink})\overline{B}_\alpha(t_{\rm source}) \rangle,
\end{equation}
with \(\displaystyle  t=t_{\rm sink}-t_{\rm source}\).
We insert an appropriate operator \(O(\vec{q}, t')\) at time \(t'\), \(t_{\rm source}\le t' \le t_{\rm sink}\), and possibly finite momentum transfer \(\vec{q}\), to obtain a form factor or structure function moment three-point correlation function,
\begin{equation}
C^{\Gamma, O}_{\rm 3pt}(t, t', \vec{q}) =
\sum_{\alpha,\beta}
\Gamma_{\alpha\beta}
\langle B_\beta(t_{\rm sink}) O(\vec{q}, t') \overline{B}_\alpha(t_{\rm source}) \rangle,
\end{equation}
with appropriate projection, \(\displaystyle \Gamma=\frac{1+\gamma_t}{2}\), for a spin-unpolarized, and \(\displaystyle \Gamma=\frac{1+\gamma_t}{2}i\gamma_5\gamma_k, k\ne 4\), for a polarized nucleon.
Ratios of these two- and three-point functions give plateaux for \(0<\tau=t'-t_{\rm source}<t\) that are fitted to a constant to extract the bare lattice matrix elements of desired observables: e.g.\ at \(q^2=0\), we use the ratio
\begin{equation}
\langle O\rangle^{\rm bare}=
\frac{C^{\Gamma, O}_{\rm 3pt}(t, \tau) }{C_{\rm 2pt}(t)}\ .
\end{equation}
In this paper we limit ourselves to those low structure function moments that are calculable at \(q^2=0\), as are listed in Table \ref{tab:OptDef}.
\begin{table}
\caption{
Operators used in the structure function moment calculations, including the notation for the operator, the explicit operator form, the hypercubic group representation, the correlator ratios and the
projection operators used in the non-perturbative renormalization of the operator in Eq.~(\ref{eq:Zri}).}
\label{tab:OptDef}
\begin{center}
\begin{tabular}{lcc}
\hline\hline
\multicolumn{2}{c}{quark momentum fraction \(\langle x \rangle_q\)} \\
\hline
\(O_\Gamma\) & \(\displaystyle {\cal O}^q_{44} = \overline{q} \left[\gamma_4
\stackrel{\leftrightarrow}{D_4} - \frac{1}{3}\sum_k \gamma_k
\stackrel{\leftrightarrow}{D_k} \right] q\) \\
hypercubic group rep.& \({\bf 3}^+_1\)\\
correlator ratio& \(\displaystyle R_{\langle x \rangle_{q}} = \frac{C^{\Gamma,{\cal O}^q_{44}}_{\rm
3pt}}{C_{\rm 2pt}} = m_N \langle x \rangle_q\)\\
NPR projection& \(
{{\cal P}^q_{44}}^{-1} =
\gamma_4 p_4 - \frac{1}{3}\sum_{i=1,3} \gamma_i p_i\)\\
\hline\hline
\multicolumn{2}{c}{quark helicity fraction \(\langle x \rangle_{\Delta q}\)}\\
\hline
\(O_\Gamma\) & \(\displaystyle {\cal O}^{5q}_{\{34\}} = i \overline{q} \gamma_5 \left[\gamma_3
\stackrel{\leftrightarrow}{D_4} + \gamma_4
\stackrel{\leftrightarrow}{D_3}  \right] q\)\\
hypercubic group rep.& \({\bf 6}^-_3\) \\
correlator ratio& \(\displaystyle R_{\langle x \rangle_{\Delta q}} = \frac{C^{\Gamma,{\cal
O}^{5q}_{\{34\}}}_{\rm 3pt}}{C_{\rm 2pt}} = m_N \langle x
\rangle_{\Delta q}\) \\
NPR projection& \(
{{\cal P}^{5q}_{34}}^{-1} = i \gamma_5 
\left( \gamma_3 p_4 + \gamma_4 p_3 \right)\)\\
\hline\hline
\multicolumn{2}{c}{transversity \(\langle 1 \rangle_{\delta q}\)}\\
\hline
\(O_\Gamma\) & \(\displaystyle {\cal O}^{\sigma q}_{34} = \overline{q} \gamma_5 \sigma_{34} q\)\\
hypercubic group rep.& \({\bf 6}^+_1\)\\
correlator ratio& \(\displaystyle R_{\langle 1 \rangle_{\delta q}} = \frac{C^{\Gamma,{\cal
O}^{\sigma q}_{\{34\}}}_{\rm 3pt}}{C_{\rm 2pt}} =
\langle 1 \rangle_{\delta q}\)\\
NPR projection& \(
{{\cal P}^{\sigma q}_{34}}^{-1} = \gamma_5 \sigma_{34}\)\\
\hline\hline
\multicolumn{2}{c}{twist-3 matrix element \(d_1\)}\\
\hline
\(O_\Gamma\) &  \(\displaystyle {\cal O}^{5q}_{[34]} = i \overline{q} \gamma_5 \left[\gamma_3
\stackrel{\leftrightarrow}{D_4} - \gamma_4
\stackrel{\leftrightarrow}{D_3}  \right] q\)
\\
hypercubic group rep.& \({\bf 6}^+_1\)\\
correlator ratio& \(\displaystyle R_{d_1} = \frac{C^{\Gamma,{\cal O}^{5q}_{[34]}}_{\rm 3pt}}{C_{\rm 2pt}} = d_1\)\\
NPR projection& \({{\cal P}^{5q}_{[34]}}^{-1} = i \gamma_5 
\left( \gamma_3 p_4 - \gamma_4 p_3 \right)\)\\
\hline\hline
\end{tabular}
\end{center}
\end{table}

The structure function moments are renormalized non-perturbatively using the Rome-Southampton regularization independent (RI-MOM) scheme \cite{Martinelli:1994ty,Dawson:1997ic}.
The chiral symmetry of DWF is relied on to suppress mixing with lattice-artifact operators.
The specific procedures for the operators studied in this paper have been described in previous RBC publications~\cite{Orginos:2005uy,Lin:2008uz}.
Here we summarize them for the readers' convenience.
\color{black}
First, we take the Fourier transform of the Green's function for operator  \(O_\Gamma\) constructed from a point source propagator at the origin,
\begin{eqnarray}
G_{O_\Gamma}(p,p';a) &=& \sum_{x,y} e^{-i p \cdot x+i p' \cdot y }\langle
\psi(x) {O_\Gamma}(0) \overline{\psi}(y) \rangle,\\
&=&  \sum_{x,y} e^{-i p \cdot x+i p' \cdot y }\langle
 S(x,0){\Gamma}S(0,y) \rangle,
\label{eq:greenFuc}
\end{eqnarray}
where \(O_\Gamma\) is  one of \({\cal O}^q_{44}\), \({\cal
O}^{\sigma q}_{34}\), \({\cal O}^{5q}_{\{34\}}\) or \({\cal
O}^{5q}_{[34]}\). The needed Fourier-transformed point-source and point-split--source propagators are
\begin{eqnarray}
S(p;a)&=&\sum_x e^{-ip\cdot x}S(x;0)\\
D_{\mu}S(p;a)&=&\sum_x \frac{1}{2}e^{-ip\cdot
x}\left[S(x;-\hat{\mu})U_\mu(-\hat{\mu})\right. \nonumber \\ &-&
\left. S(x;\hat{\mu})U_\mu^\dagger(0)\right].
\end{eqnarray}

Next the Green's function is amputated, and evaluated for the case of
exceptional momenta ($p=p'$),
\begin{eqnarray}
\Lambda_{O_\Gamma}(p;a) = \langle S(p;a)^{-1} \rangle G_{O_\Gamma}(p;p;a)\langle  S(p;a)^{-1}\rangle.
\end{eqnarray}
\(Z^{\rm RI}\) is obtained by requiring the renormalized Green's function, after some suitable projection, be equal to its tree-level counterpart~\cite{Martinelli:1994ty},
\begin{eqnarray}\label{eq:Zri}
Z_{O_\Gamma}(\mu; a)^{-1} Z_q(\mu; a)  = \frac{1}{12}{{\rm Tr}\left(\Lambda_{O_{\Gamma}}(p;a)
  P_\Gamma\right)}|_{p^2=\mu^2}.
\end{eqnarray}
The projectors for each \(O_\Gamma\) are listed in Table~\ref{tab:OptDef}. 
Since we wish to match these renormalized operators to the perturbative $\overline{\rm MS}$ scheme,
the renormalization scale \(\mu\) must be large enough for perturbation theory to be valid, but not so large to introduce lattice artifacts.   Thus $\mu$ should satisfy 
\(\Lambda_{\rm QCD} \ll \mu \ll1/a\). In practice, we have found that the upper bound is not so strict, and can be replaced by the milder condition that $(pa)^{2}<3$.

Finally, the following steps, similar to those from~\cite{Lin:2008uz}, allow us to convert the renormalization constants to the continuum \(\overline{\rm MS}\) scheme at 2~GeV.
\begin{enumerate}
\item Obtain \(Z^{\rm RI}(\mu)\): The ratio of \(Z_{O_\Gamma}(\mu; a)/Z_q(\mu; a)\) to \(Z_A/Z_q(\mu; a)\) is computed and yields  \(Z_{O_\Gamma}(\mu; a)/Z_A\). Each of the factors in the ratio is first extrapolated to the chiral limit, \(m_f=-m_{\rm res}\), at fixed momentum.
Using $Z_A = 0.7161$~\cite{Allton:2008pn},
we can determine \(Z_{O_\Gamma}(\mu; a)\) in Eq.~(\ref{eq:Zri}), the renormalization constant in the RI scheme, which we denote as \(Z^{\rm RI}\).
\item Convert to \(\overline{\rm MS}\) scheme: We are interested in continuum quantities, mostly calculated in the \(\overline{\rm MS}\) scheme.
The conversion factors between RI and \(\overline{\rm MS}\) schemes for the operators discussed here
have been calculated in Refs.~\cite{Gockeler:1998ye,Floratos:1977au}.
To get \(Z^{\rm\overline{MS}}(\mu)\), 
we use $\alpha_s^{(3)}(\mu)$ obtained by numerically solving the
     renormalization group equation with the four-loop anomalous dimension \cite{vanRitbergen:1997va}
     and initial condition $\alpha_s^{(5)}(m_Z)=0.1176$
     \cite{Yao:2006px}, following the method in Appendix A of
     Ref.~\cite{Aoki:2007xm}.
%
\item Running to 2 GeV with two loop anomalous dimensions
     [31] \cite{Broadhurst:1994se}. This will take away the
     continuum running factor.
\item 
Remove $(ap)^2$ lattice artifacts: We fit the remaining momentum
     dependence, which we will observe to be very small, to the
     form $f = A(ap)^2+B$ and finally get $Z^{\overline{\rm MS}}(2 \mbox{GeV})$.
\end{enumerate}

\section{Ensembles}
\label{sec:Ensembles}

We follow the same sampling procedure as in our nucleon form factor calculations reported in  Ref.~\cite{Yamazaki:2009zq}.

\subsection{Statistics}

The RBC-UKQCD joint (2+1)-flavor dynamical DWF ensembles \cite{Allton:2008pn} are used for the calculations.
These ensembles are generated with Iwasaki gauge action~\cite{Iwasaki:1983yi} at the coupling \(\beta=2.13\) which corresponds to
the lattice cutoff of \(a^{-1}=1.73(3)\) GeV, which is determined
from the \(\Omega^{-}\) baryon mass~\cite{Allton:2008pn}.

The dynamical up, down and strange quarks are described by DWF actions with fifth-dimensional extent of \(L_s=16\) and the domain-wall height of \(M_5=1.8\).
The strange quark mass is set at 0.04 in lattice units and turned out
to be about 12\% heavier than the physical
strange quark, after taking into account the additive correction of the residual mass, \(m_{\rm res}=0.00315\).
The degenerate light quark masses in lattice units, 0.005, 0.01, 0.02 and 0.03, correspond to pion masses of about 0.33, 0.42, 0.56 and 0.67 GeV and nucleon masses, 1.15, 1.22, 1.38 and 1.55 GeV.

Two lattice sizes are used for our study, \(16^3\times 32\) and \(24^3\times 64\), corresponding to linear spatial extent of approximately 1.8 and 2.7 fm, respectively.
The smaller volume ensembles, calculated only with the heavier three light quark masses, are used for a finite-volume study. 
On the \(16^3\) ensembles we use 3500 trajectories separated by 5 trajectories at \(m_f = 0.01\) and 0.02, and by 10 at 0.03.
The main results are obtained from the larger volume ensembles with the number of the configurations summarized in Table~\ref{table:statistics}.
\begin{table}[!t]
\caption{\(N_{\rm conf}\), \(N_{\rm sep}\) and \(N_{\rm meas}\) denote number of gauge configurations, trajectory separation between measurements, and the number of measurements on each configuration, respectively, on the 24\(^3\) ensembles. The table also lists the pion and nucleon mass for each ensemble~\cite{Yamazaki:2009zq}.}
\begin{center}
\begin{tabular}{lr@{}lcr@{}lll} \hline\hline
\multicolumn{1}{l}{\(m_f\)} & 
\multicolumn{2}{l}{\(N_{\rm conf}\)} &
\multicolumn{1}{l}{\(N_{\rm sep}\)} &
\multicolumn{2}{l}{\(N_{\rm meas}\)} &
\multicolumn{1}{l}{\(m_\pi\)[GeV]} &
\multicolumn{1}{l}{\(M_N\)[GeV]}\\ \hline
0.005 \hfill& 932 &\footnotemark[1]& 10\hfill & \hfill 4 &\footnotemark[2] & 0.3294(13) &1.154(7)\\ 
0.01   & 356 & & 10 & 4 & & 0.4164(12) &1.216(7)\\
0.02   &  98 & & 20 & 4 & & 0.5550(12) &1.381(12)\\
0.03  & 106 & & 20 & 4 & & 0.6681(15) &1.546(12)\\ \hline\hline
\end{tabular}
\end{center}
\protect\footnotetext[1]{
Total number of configurations is actually 646.
We carry out extra measurements on a subset of these (286 configurations) to improve the statistics using different source positions.}
\protect\footnotetext[2]{
Two measurements with the double-source method gives effectively four measurements.}
\label{table:statistics}
\end{table}

On the larger volume, at the heavier three quark masses we make four measurements on each configuration with the conventional single source method using \(t_{\rm src} = 0,\) 16, 32, 48, or 8, 19, 40, 51.
At the lightest mass the double-source method \cite{Yamazaki:2009zq} is used,
and two measurements on each configuration are carried out using the source pairs of (0, 32) and (16, 48), or (8, 40) and (19, 51).
We made an additional two measurements on roughly half of the configurations with one or the other source pair. This means that 
we make four, double-source measurements on almost half of the configurations, while two double-source measurements are carried out on the remaining configurations.

On the smaller volume, a single source is used, however the location of this source is shifted for each successive measurement in the order (x,y,z,t)=(0,0,0,0), (4,4,4,8), (8,8,8,16), (12,12,12,24), reducing autocorrelations.

In order to reduce possible auto-correlations among measurements, 
they are averaged on each configuration and then 
blocked into bins of 40 trajectories for the $24^{3}$ ensembles, and 20 trajectories for the $16^{3}$ ensembles. The statistical errors are estimated by the jackknife method on the blocked measurements.

Finally, the non-perturbative renormalization constants were computed on the four 24$^{3}$ ensembles, on roughly 50 configurations each, separated by 40 trajectories. The maximum momentum value in units of $2\pi/L_i$ in each direction was 6 (spatial) and 17 (temporal), such that $(pa)^2<3$.

\subsection{Correlation functions}

The quark propagator is calculated with an anti-periodic boundary condition in the temporal direction, and periodic boundary conditions for the spatial directions.
We employ gauge-invariant Gaussian smearing~\cite{Alexandrou:1992ti, Berruto:2005hg} at the source with smearing parameters \((N,\omega) = (100,7)\) which were chosen after a series of pilot calculations.
For the calculation of the three-point functions, we use a time separation of 12 time slices between the source and sink operators to reduce effects from excited state contributions as much as possible.

\subsection{Systematic errors}

There are several important sources of systematic error that need be considered: finite spatial size of the lattice, excited state contamination, and non-zero lattice spacing.
A chiral-perturbation-theory-inspired analysis of the former 
for meson observables suggests the dimensionless product, \(m_\pi L\), of the
calculated pion mass \(m_\pi\) and lattice linear spatial extent
\(L\), should be set greater than 4 to ensure that the finite-volume correction is negligible (below one percent), and the available lattice calculations seem to support this.
While our present parameters satisfy this condition, we discovered that even our larger volume of \((\mbox{\rm 2.7 fm})^3\) is insufficient for calculating such important nucleon properties as the axial charge \cite{Yamazaki:2008py} and form factors \cite{Yamazaki:2009zq}:
As we reduce the light quark mass, 
eventually to \(m_\pi = 330\) MeV and \(m_\pi L \sim 4\),
finite-size effects become severe,
exceeding 10 \%,
at least for these quantities that are measured in elastic processes.
Similar finite-size effects may influence the moments of structure functions we are discussing in this paper (studies of finite volume effects in chiral perturbation theory can be found in \cite{Detmold:2003rq,Detmold:2005xj}).
On the other hand since these moments are extracted from experimental observables measured in very different inelastic processes, the finite-size effect may enter differently.
It is an important goal of this work to investigate such finite-size effects on the moments of structure functions.

\begin{figure}
\begin{center}
\includegraphics[width=\columnwidth,clip]{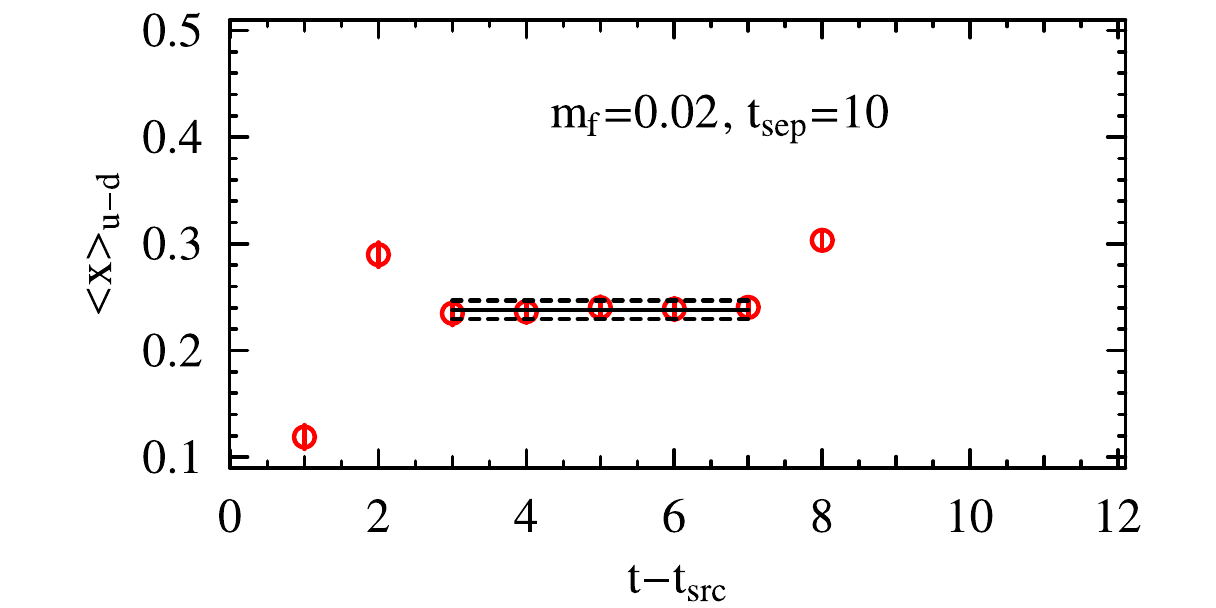}
\includegraphics[width=\columnwidth,clip]{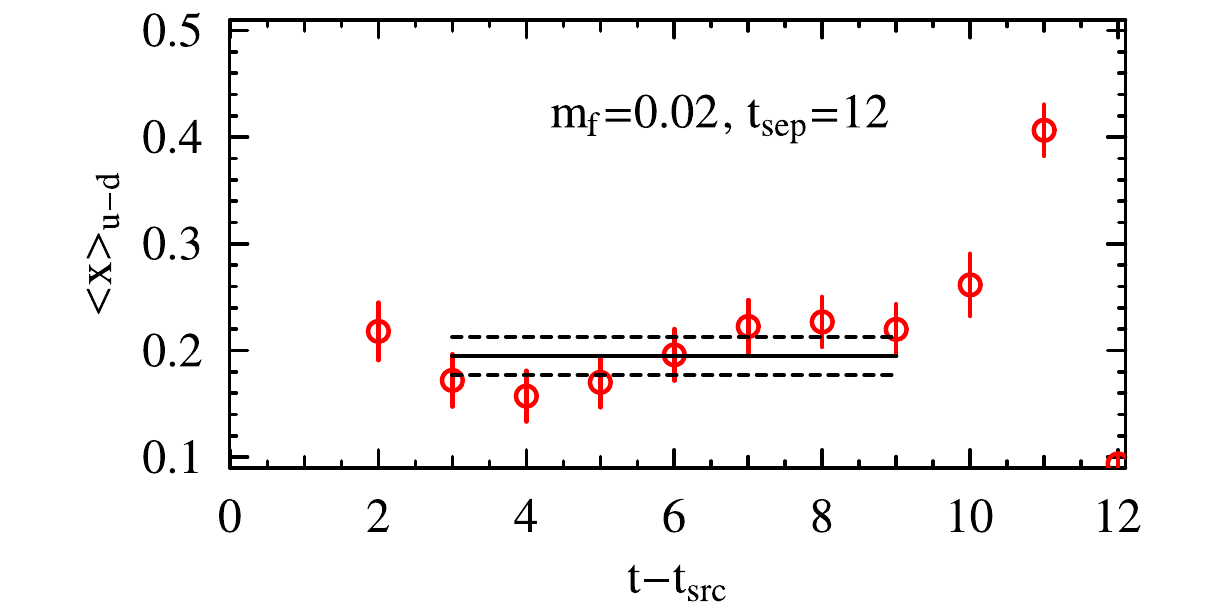}
\end{center}
\caption{A nucleon observable, isovector quark momentum fraction, \(\langle x\rangle_{u-d}\), from RBC 2-flavor dynamical DWF ensemble with \(m_{ud}=0.02\) \cite{Lin:2008uz}, with source-sink separation of 10 and 12: a clear systematic difference is seen.  The shorter source-sink separation is not manifestly free of excited-state contamination.\label{fig:source-sink}}
\end{figure}

\begin{figure}[!th]
\begin{center}
\includegraphics[width=\columnwidth,clip]{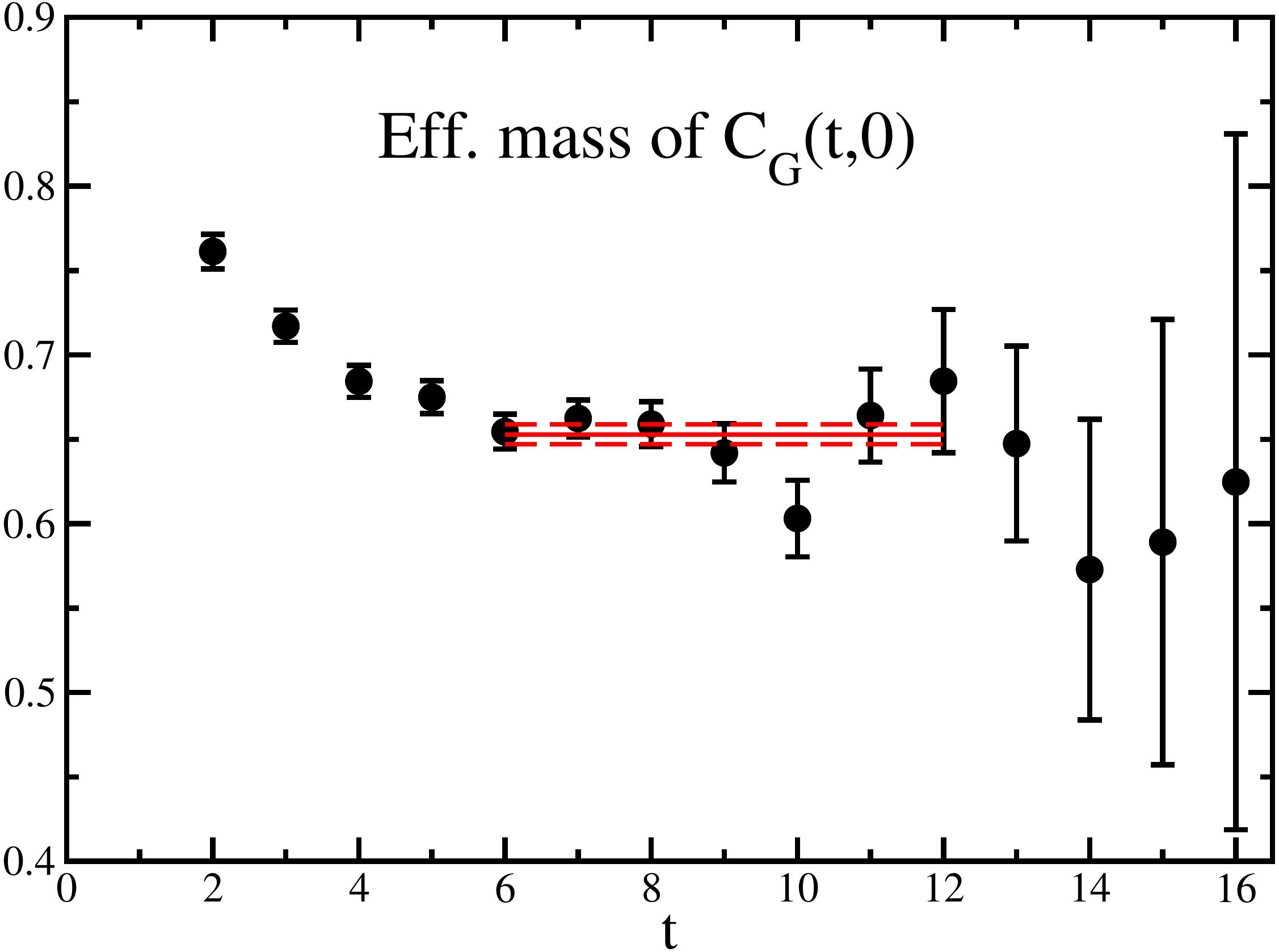}
\end{center}
\caption{Effective mass of the nucleon correlation function with Gaussian smearing
applied at both source and sink, for quark mass \(m_f=0.005\).
\label{fig:Nemass}
}
\end{figure}

In order to reduce contamination from excited states, it is important to adjust the time separation between the nucleon source and sink appropriately so the resultant nucleon observables are free of contamination from excited states.
The separation has to be made longer as the quark mass is decreased.
In our previous study with two dynamical flavors of DWF quarks~\cite{Lin:2008uz} with a similar lattice cutoff of about 1.7 GeV, we saw systematic differences between observables calculated with the shorter time separation of 10, or about 1.16 fm, and longer 12, or 1.39 fm: the differences amount to about 20 \%, or two standard deviations (see Fig.\ \ref{fig:source-sink}.)
This would suggest that at the shorter time separation of about 1.2 fm, the excited-state contamination has not decayed sufficiently to guarantee correct calculations for the ground-state observables~\cite{Ohta:2008kd}.
\color{black}
While it is desirable to use a longer separation, it cannot be made too long in practice without losing control of statistical errors.
In Fig. \ref{fig:Nemass} we present the nucleon effective mass at the lightest quark mass, \(m_f=0.005\).
The nucleon signal begins to decay at \(t=12\), or about 1.4 fm: this is about longest distance we can choose without losing the signal.
As will be shown in detail in this paper, the bare three-point function signals for this source-sink separation of \(t=12\) are acceptable. 
We note that recently the LHP Collaboration has also looked at this issue in some detail~\cite{Bratt:2010jn} and ends up using a shorter separation of about 1.2 fm.

For low energy quantities like the pseudoscalar decay constants, the kaon $B$-parameter, and the $\Omega$ baryon mass, the effect of non-zero lattice spacing was estimated to be less than 4\% for the configuration ensemble used in this work~\cite{Allton:2008pn}, and subsequently confirmed on a later ensemble with smaller lattice spacing~\cite{Kelly:2009fp,Mawhinney:2009jy}. We expect that similar errors hold for the quantities discussed in this paper. 
\section{Numerical results}
\label{sec:fractions}

\subsection{Quark momentum and helicity fractions}
\label{sec:fraction_xq_xdq}

Let us first discuss the ratio, \(\langle x \rangle_{u-d}/\langle x \rangle_{\Delta u - \Delta d}\), of the isovector quark momentum fraction to the helicity fraction.
The momentum fraction, \(\langle x \rangle_{u-d}\), which is the first moment of the \(F_{1,2}\) unpolarized structure functions, and the helicity fraction, \(\langle x \rangle_{\Delta u - \Delta d}\), which is the first moment of the \(g_1\) polarized structure function, share a common renormalization because they are related by a chiral rotation and the DWF action preserves chiral symmetry to a high degree.
Thus, this ratio calculated on the lattice is naturally renormalized, much like the form factor ratio \cite{Yamazaki:2009zq}, \(g_A/g_V\), and is directly comparable with the value obtained from experiment.

\begin{figure}
\begin{center}
\includegraphics[width=\columnwidth,clip]{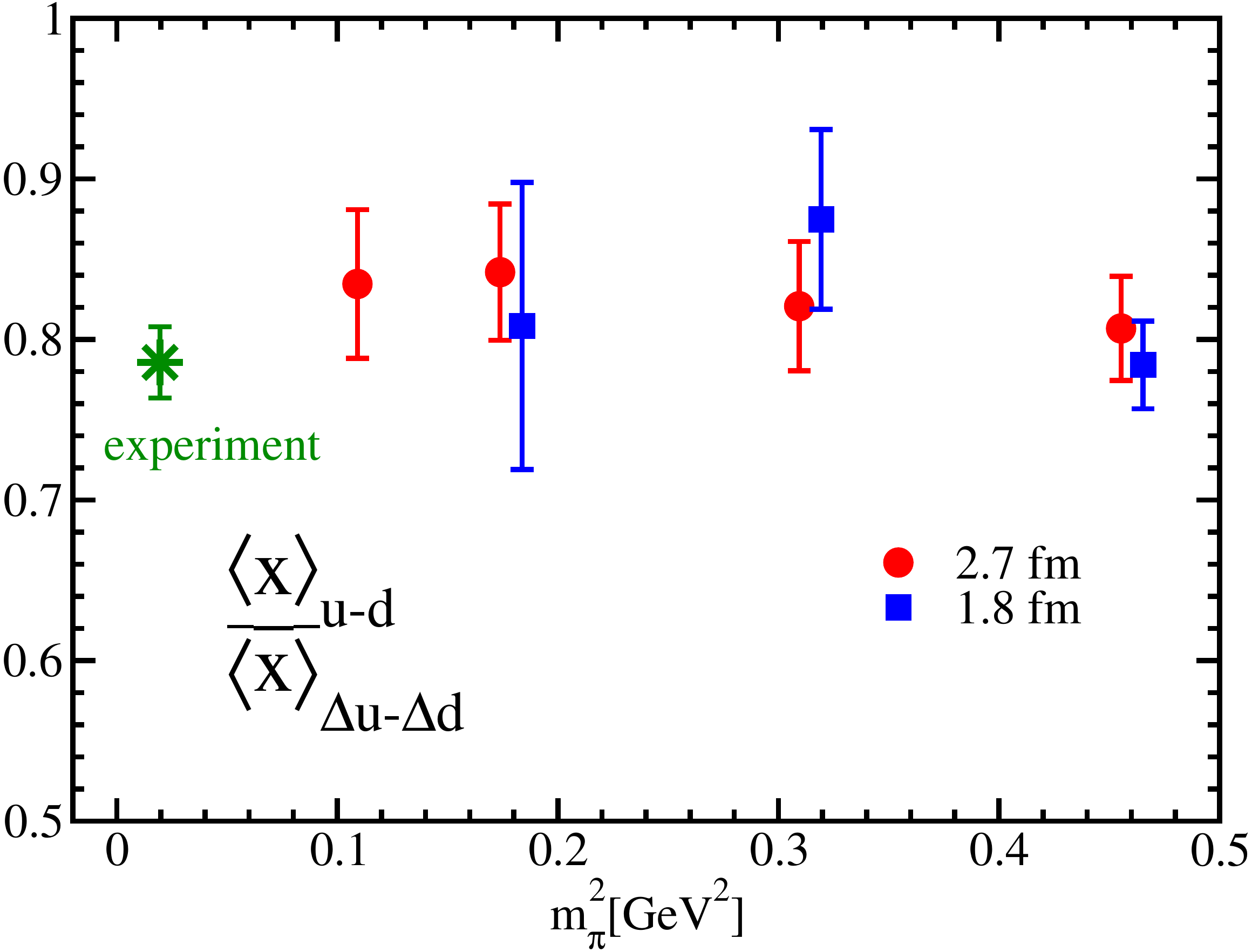}
\end{center}
\caption{
Ratio of the bare, isovector, momentum and helicity fractions, \(\langle x\rangle_{u-d}/\langle x\rangle_{\Delta u-\Delta d}\), which is naturally renormalized for DWF. Both volumes are shown, \(({\rm 2.7\  fm})^3\) (circles) and \(({\rm 1.8\ fm})^3\) (squares).
The square symbols have been moved slightly in the plus x-direction.
They are in good agreement with experiment which is denoted by the star.
No discernible dependence on volume nor pion mass can be detected.
\label{fig:xqxDqratio}}
\end{figure}

The results of our calculation are shown in Fig. \ref{fig:xqxDqratio}.
They do not show any discernible dependence on the up/down quark mass, outside of the statistical error bars, and are in good agreement with experiment.
This is in contrast to the renormalized ratio of \(g_A/g_V\) of elastic form factors which at the lightest point deviates significantly from heavier mass results and the experiment as a result of a large finite-size effect~\cite{Yamazaki:2009zq}.
This suggests the moments of inelastic structure functions such as the momentum fraction, \(\langle x \rangle_{u-d}\), and helicity fraction, \(\langle x \rangle_{\Delta u - \Delta d}\), may not suffer so severely from the finite-size effect that plagues elastic form factor calculations.
Indeed the results obtained from the smaller \( (1.8\,{\rm fm})^3 \) volume, also shown 
in Fig.~ \ref{fig:xqxDqratio}, do not deviate significantly from the constant behavior of the larger volume results,  albeit with larger statistical errors.

\begin{figure}
\begin{center}
\includegraphics[width=1.2\columnwidth,clip]{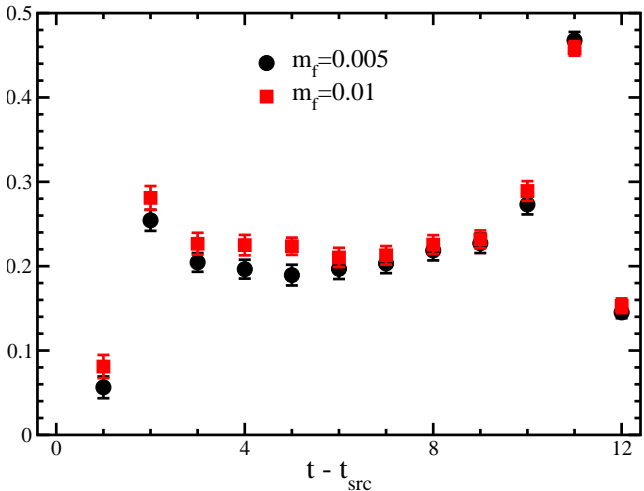}
\end{center}
\vskip -.5cm
\caption{Signals for the ratio of three- and two-point functions for the bare quark momentum fraction, \(\langle x\rangle_{u-d}\). Quark mass \(0.005\) (circles) and 0.01 (squares).
\label{fig:xqsignal}}
\end{figure}

Next we discuss the absolute values of the isovector quark momentum fraction, 
\(\langle x \rangle_{u-d}\).
This is the first moment of the unpolarized structure functions, \(F_1\) and \(F_2\).
In Fig. \ref{fig:xqsignal}, we show the bare lattice matrix elements as ratios of three-
and two-point functions for the two lightest quark mass values of \( m_f=0.005\) (circles)
and 0.01 (squares). 
We extract bare values of the desired matrix element by averaging over time slices 4 to 8 (values are summarized in Tables \ref{tab:Xud_L24} and \ref{tab:Xud_L16}).

\begin{figure}
\begin{center}
\includegraphics[width=\columnwidth,clip]{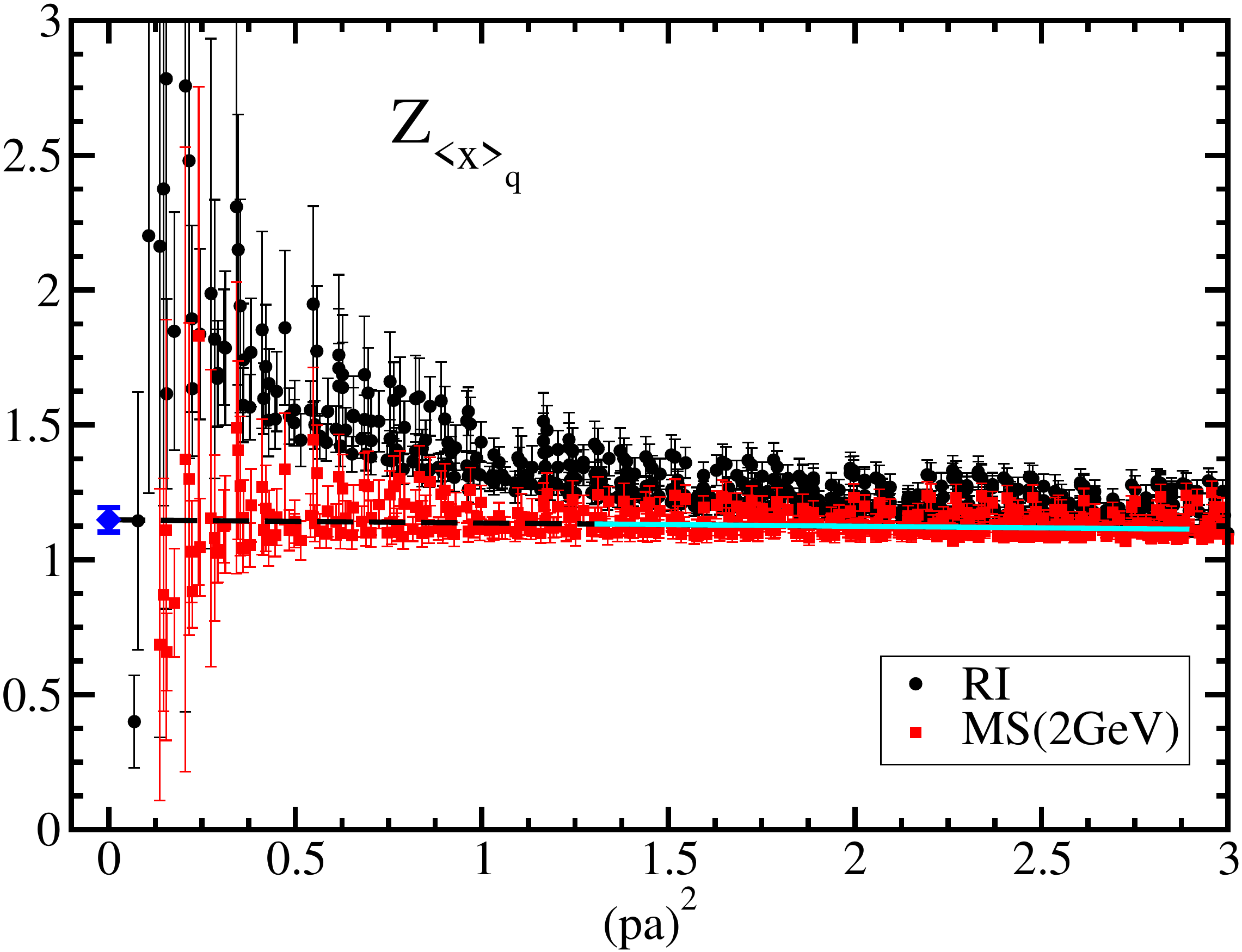}
\end{center}
\caption{Non-perturbative renormalization factor for the quark momentum fraction, \(\langle x \rangle_{u-d}\)
. Circles denote the RI-MOM values, squares the $\overline{\rm MS}$ ones. The line denotes a linear fit used to remove the leading $O((ap)^2)$ lattice artifacts.
\label{fig:xqnpr}}
\end{figure}

These bare values need be renormalized in order to be compared with experiment.
In Fig.\ \ref{fig:xqnpr} we present the non-perturbatively determined renormalization for
the operator \({\cal O}_{44}^{q}\).
The filled circles are the renormalization constants in the RI-MOM scheme at scale $\mu^2=p^2$, which
is not scale independent. The filled squares correspond to the renormalisation constant given in the \(\overline{\rm MS}\) scheme at $\mu=2$ GeV, where there remains only residual scale dependence proportional to \(  (ap)^2 \) lattice artifacts. After removing the remaining \((ap)^2\) dependence as 
described in Sec.~\ref{sec:Formulation},
we obtain a renormalization factor of  \(Z^{\overline{\rm MS}}_{\langle x\rangle_{q}}({\rm 2~GeV}) = 1.15(4)\).

\begin{figure}
\begin{center}
\includegraphics[width=\columnwidth,clip]{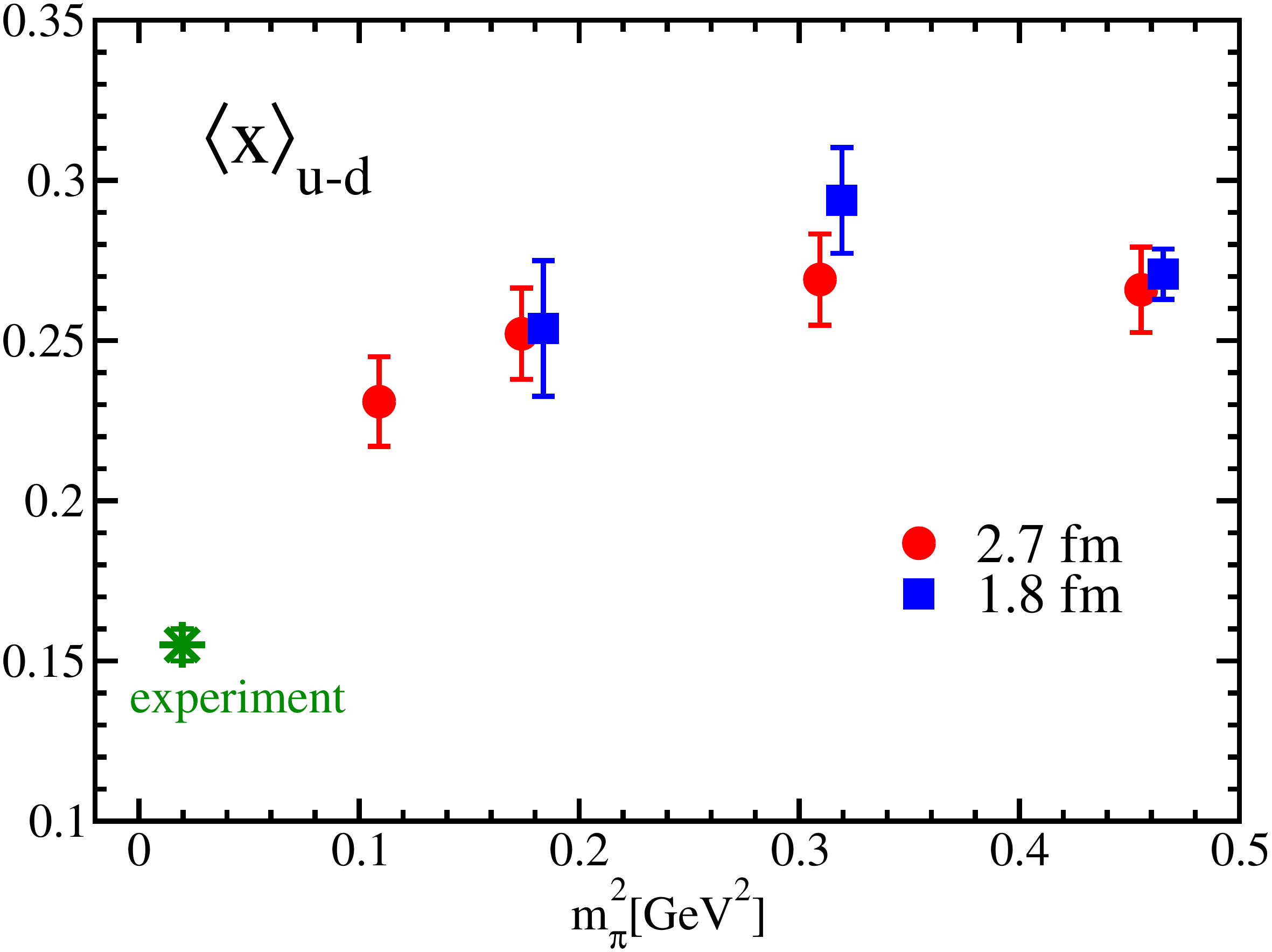}
\end{center}
\caption{Renormalized quark momentum fraction, \(\langle x \rangle_{u-d}\). Both volumes are shown, \(({\rm 2.7\  fm})^3\) (circles) and \(({\rm 1.8\ fm})^3\) (squares).
The square symbols have been moved slightly in the plus x-direction for clarity.
\label{fig:xq}}
\end{figure}

Using this renormalization constant, the quark-mass dependence of the momentum fraction 
is shown in Fig.\ \ref{fig:xq}.
The results from the smaller (1.8-fm)$^3$ volume for the heavier three quark mass values are in agreement with respective mass-value results from the larger volume.
These heavier points stay roughly the constant which is about 70 \% higher at \(\sim 0.26\) 
than the experiment, about 0.15. This behavior is not so different from old RBC quenched results \cite{Orginos:2005uy} and other recent ones~\cite{Renner:2010ks} with similar up/down quark mass.

On the other hand, the lightest point on the larger (2.7 fm)$^3$ volume, shows a sign of deviation away from this constant behavior. In contrast to the form factor deviations that move away from experiment \cite{Yamazaki:2009zq},
this one trends toward the experimental value.
Since a lighter quark can more easily share its momentum with other degrees of freedom, this trending toward the experiment may well be a real physical effect:
It is not necessarily a result of the finite spatial size of the lattice.

Indeed, it is noteworthy that the \(m_\pi L\) value of 3.8 for \(m_f=0.01\) at \(L=1.8\) fm is smaller than that of 4.5 for \(m_f=0.005\) at \(L=2.7\) fm.
In other words, if  there would be such a finite-size effect for this quark momentum fraction that scales with \(m_\pi L\) as seen in the form factors, the result from  \(m_f=0.01\) at \(L=1.8\) fm should move away from that of \(m_f=0.005\) at \(L=2.7\) fm. We note that in \cite{Detmold:2003rq} it was predicted that finite volume effects in $\langle x \rangle_{u-d}$ would only become noticeable for very light quark masses. 
\color{black}

\begin{figure}
\begin{center}

\includegraphics[width=1.2\columnwidth,clip]{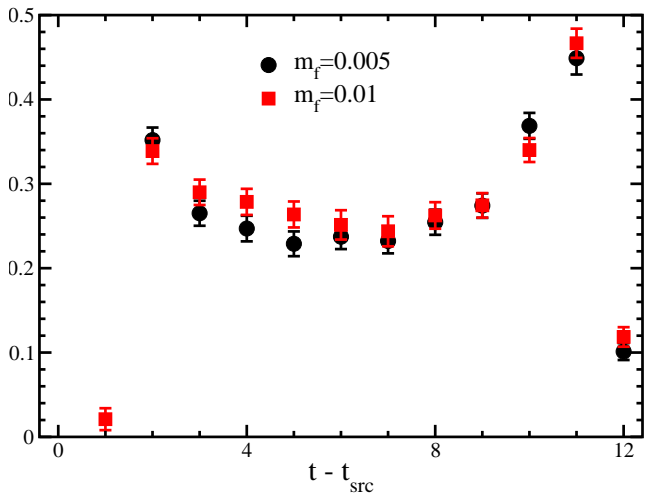}
\end{center}
\vskip -.5cm
\caption{
Signals for the ratio of three- and two-point functions for the bare quark helicity fraction, \(\langle x\rangle_{\Delta u-\Delta d}\). Quark mass \(0.005\) (circles) and 0.01 (squares).
\label{fig:xDqsignal}}
\end{figure}

The isovector quark helicity fraction, \(\langle x \rangle_{\Delta u - \Delta d}\), appears as a leading twist moment of the polarized structure functions \(g_1\) and \(g_2\).
Figure \ref{fig:xDqsignal} presents typical bare signals of this quantity on the larger (2.7-fm)$^3$ volume, for the light-quark mass points \(m_f=0.005\) and 0.01.
Average values are extracted as for the momentum fraction, described above.
Results are summarized in Tables \ref{tab:Xud_L24} for the larger volume
and Table \ref{tab:Xud_L16} for the smaller volume.

\begin{figure}
\begin{center}
\includegraphics[width=\columnwidth,clip]{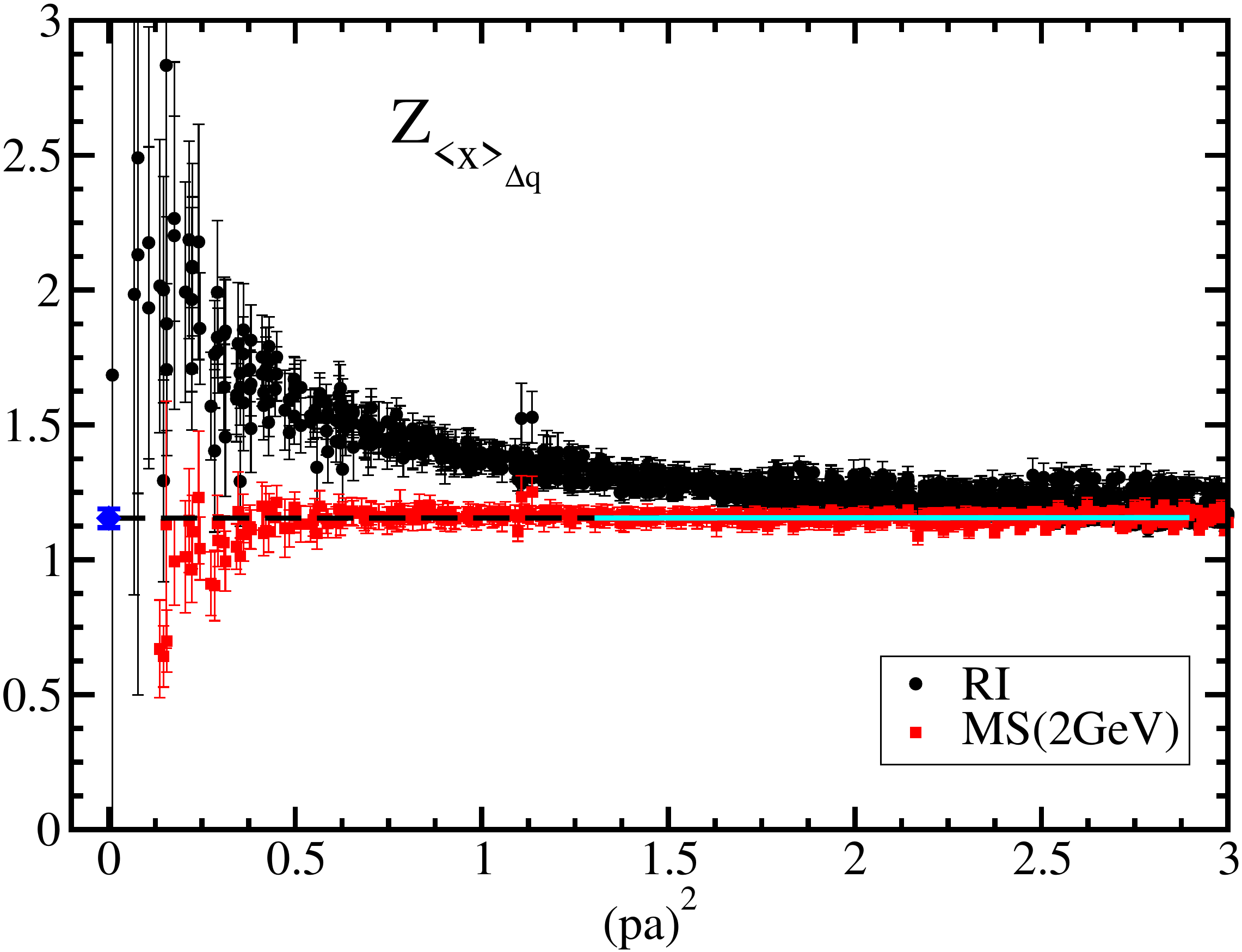}
\end{center}
\caption{Non-perturbative renormalization for the quark helicity fraction, \(\langle x \rangle_{\Delta u-\Delta d}\). Circles denote the RI-MOM values, squares the $\overline{\rm MS}$ ones. The line denotes a linear fit used to remove the leading $O((ap)^2)$ lattice artifacts.
\label{fig:xDqnpr}}
\end{figure}

The non-perturbative renormalization of \(\langle x \rangle_{\Delta u - \Delta d}\) (for the operator \({\cal O}_{\{34\}}^{5q}\)), is presented in Fig.\ \ref{fig:xDqnpr}. 
We obtain a renormalization factor of  \(Z^{\overline{\rm MS}}_{\langle x\rangle_{\Delta q}}({\rm 2 GeV}) = 1.15(3)\) through the same procedure described previously.
This value agrees very well with the corresponding value for the momentum fraction, as guaranteed by the chiral symmetry of DWF, justifying our use of bare quantities in the ratio, \(\langle x\rangle_{u-d}/\langle x\rangle_{\Delta u-\Delta d}\), earlier in this section.
\color{black}

\begin{figure}
\begin{center}
\includegraphics[width=\columnwidth,clip]{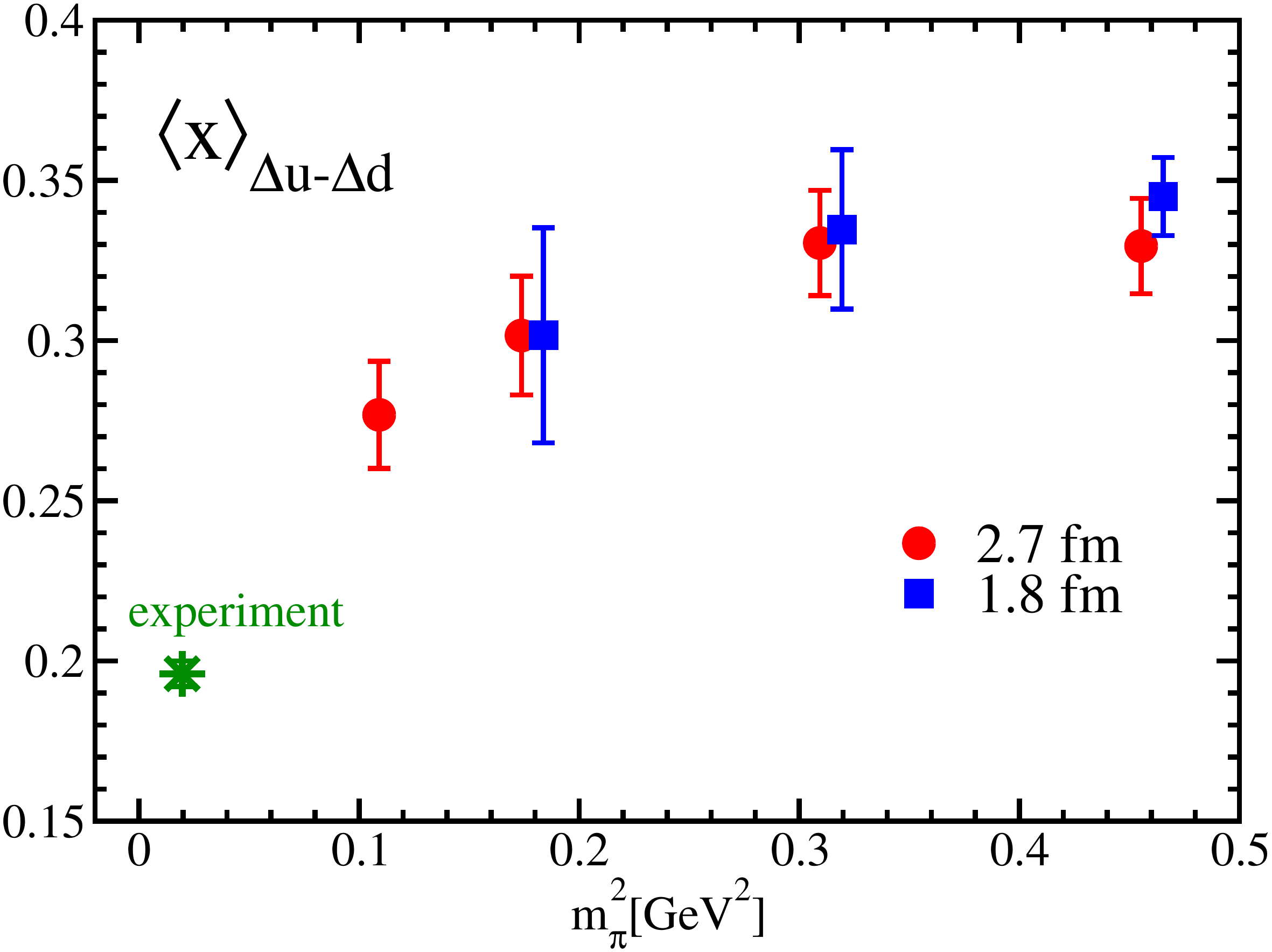}
\end{center}
\caption{The renormalized quark helicity fraction, \(\langle x \rangle_{\Delta u - \Delta d}\). 
Both volumes are shown, \(({\rm 2.7\  fm})^3\) (circles) and \(({\rm 1.8\ fm})^3\) (squares).
The square symbols have been moved slightly in the plus x-direction for clarity.
\label{fig:xDq}}
\end{figure}

With this renormalization, \(\langle x \rangle_{\Delta u - \Delta d}\) can be compared with the experiment (see Fig.\ \ref{fig:xDq}.)
No finite volume effect is apparent in the data, similar to the quark 
momentum fraction. The three heavier-mass results from the smaller volume again agree with the respective larger-mass results, suggesting the huge finite-size effect, seen in the elastic form factors, that appears to scale with \(m_\pi L\), is not present in this moment of this deep-inelastic structure function, at least at the quark masses considered here.

Moreover, the observable exhibits very similar quark-mass dependence 
to the momentum fraction, as can be expected from the near constant behavior of their ratio:
the three heavier points stay roughly the constant and about 70 \% higher than 
the experimental value, and the lightest point shows a sign of deviation away from this 
constant behavior. This trend toward the experimental value may be a real physical effect.

Here, we note that while our results of \(\langle x\rangle_{u-d}\)
are in agreement with \(n_f=2\) Wilson results \cite{Gockeler:2009pe}, they
 differ significantly from the LHP mixed-action calculations \cite{Hagler:2007xi}.
Their values are significantly lower, by about 20 \%.
The main source of this discrepancy is likely due to the use of
perturbative renormalization by the LHP Collaboration.
In the LHP mixed-action calculations~\cite{Hagler:2007xi}, the renormalization
constant is evaluated by $Z_{\cal O}=(Z_{\cal O}/Z_A)^{\rm pert.}\times Z_A^{\rm non-pert.}$
for the operator ${\cal O}$. In the same manner, we use the value of $(Z_{\cal O}/Z_A)^{\rm pert.}$ for the operator \({\cal O}^{5q}_{\{34\}}\)  from Ref.~\cite{Boyle:2006pw} and $Z_A^{\rm non-pert.}=0.7161(1)$ from Ref.~\cite{Allton:2008pn}, and evolve it to a renormalization scale of 2 GeV using the two-loop anomalous dimension~\cite{{Floratos:1977au},{Lin:2008uz}}, obtaining the renormalization factor \( Z_{\langle x \rangle_{\Delta q}}(2{\rm GeV})=0.873(34)\) in the \(\overline{\rm {MS}}\) scheme.

\begin{figure*}
\begin{center}
\includegraphics[width=\columnwidth,clip]{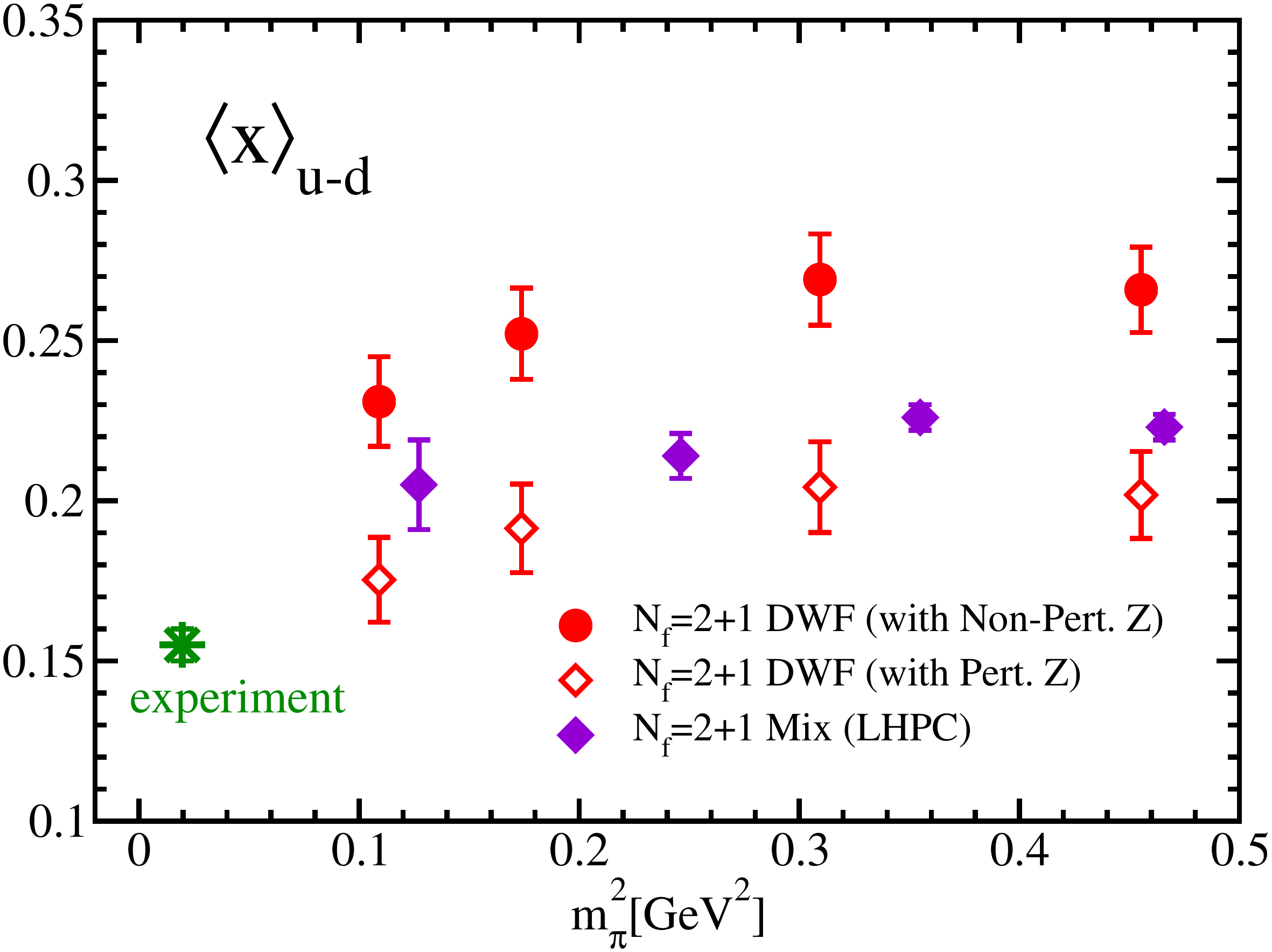}
\includegraphics[width=\columnwidth,clip]{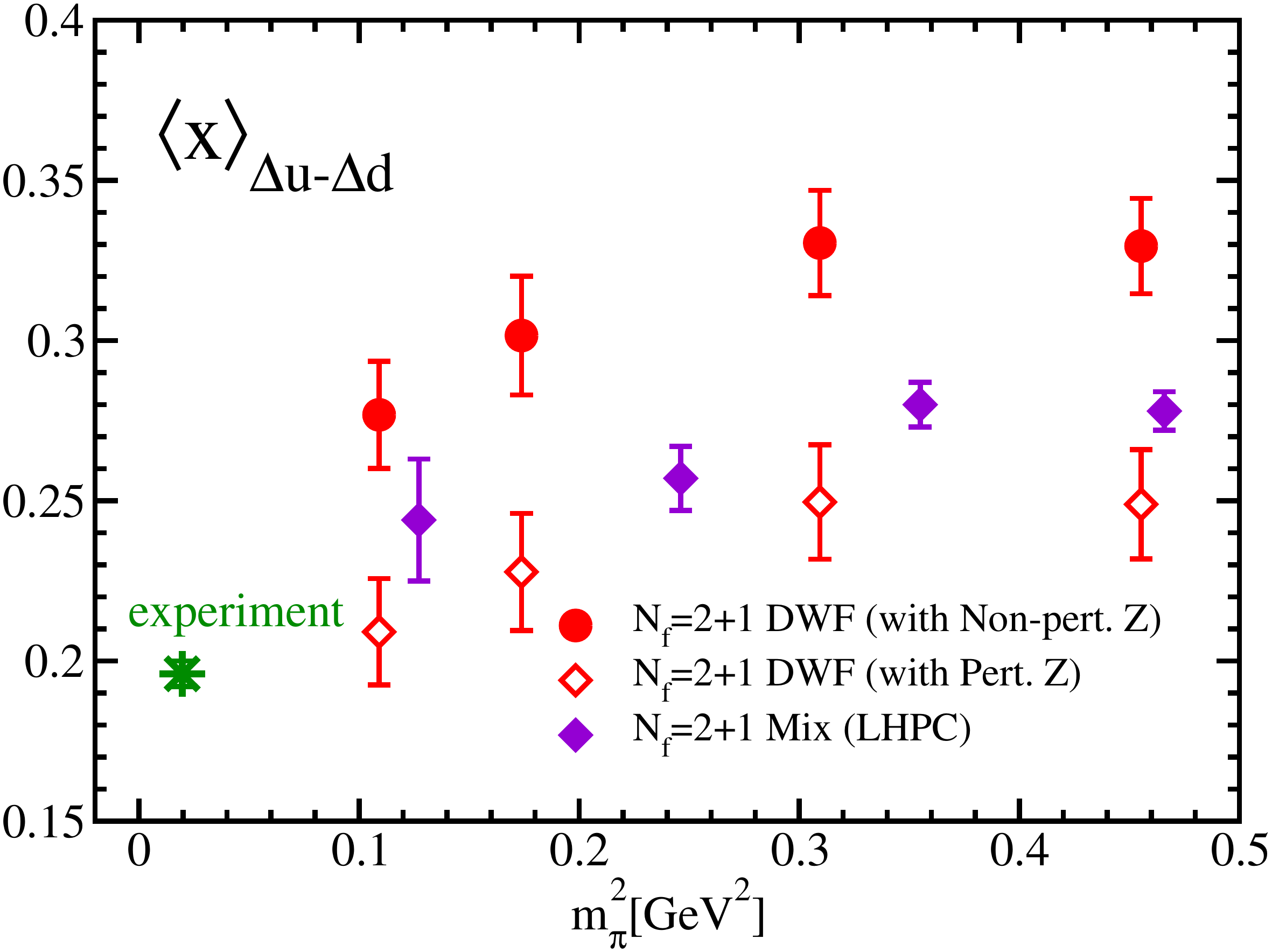}
\end{center}
\caption{Comparison with results obtained from the mixed-action calculation~\cite{Hagler:2007xi} of the LHP Collaboration (filled diamonds).
The left (right) panel is for the renormalized value of quark momentum (helicity) fraction.
Our fully non-perturbatively renormalized results are represented by filled circles, while
open diamonds denote our estimates of the same quantity with the renormalization constant determined perturbatively.}
\label{fig:mfxq_xdq_pert}
\end{figure*}

If we use this renormalization factor instead of the non-perturbative one described earlier, as shown 
in Fig.~\ref{fig:mfxq_xdq_pert}, our results will be consistent with the LHPC results. The 
difference between the non-perturbative and perturbative renormalization factors suggests a systematic error of about 25\% should be assigned to the latter.
Furthermore, those lightest points in 
both quark momentum and helicity fractions are quite close to the experiments, while the non-perturbatively renormalized ones are significantly away from the experiments. This indicates 
that the perturbative calculation of the renormalization constants significantly 
underestimates the renormalized value of these particular quantities, and then exhibits
an accidental consistency with the experiments. 

\begin{figure*}
\begin{center}
\includegraphics[width=\columnwidth,clip]{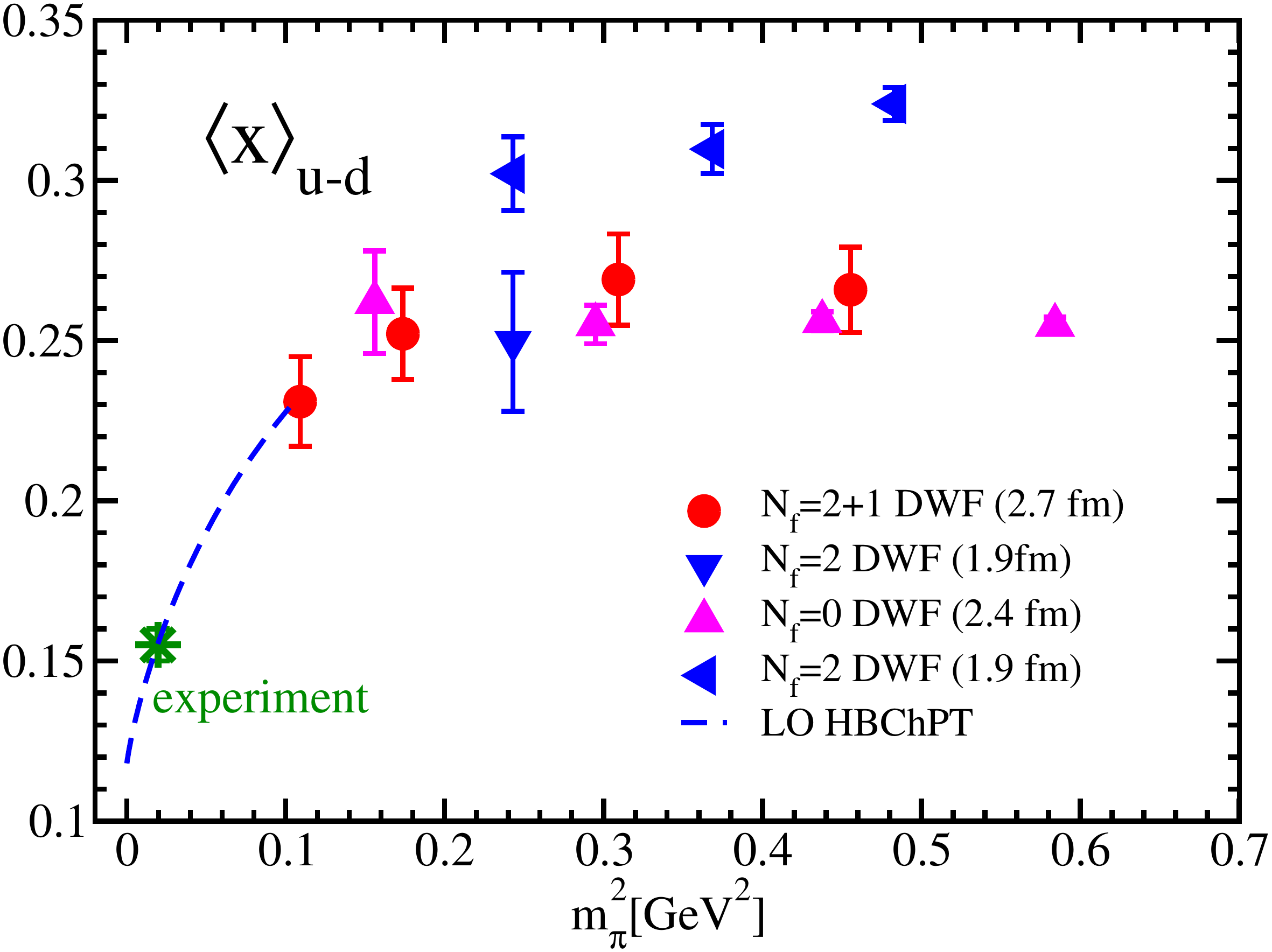}
\includegraphics[width=\columnwidth,clip]{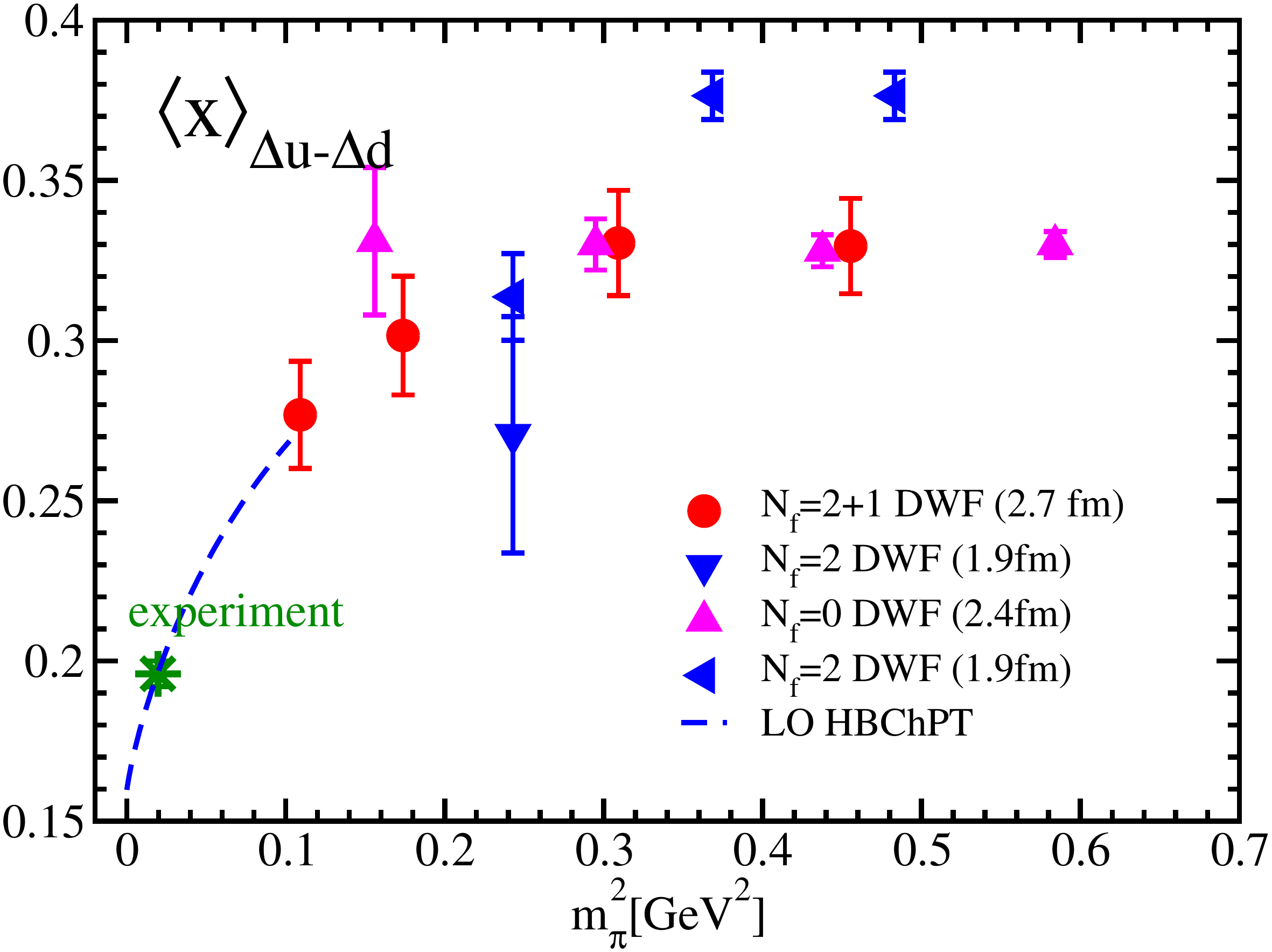}
\end{center}
\caption{
Comparison with previous RBC quenched~\cite{Orginos:2005uy} and 2-flavor~\cite{Lin:2008uz} results.  
The left triangles denote 2-flavor points with a time separation between source and sink of 12 sites instead of 10 (down triangles).
Dashed curves show the leading order behavior in HBChPT in the vicinity of the physical pion mass point.
}
\label{fig:LOChPT_xq_xdq}
\end{figure*}

As mentioned before, it is observed that there is a noticeable nonlinearity 
in the data of both \(\langle x\rangle_{u-d}\) and \(\langle x\rangle_{\Delta u-\Delta d}\).
These trends toward the experimental values are easily seen in Fig.~\ref{fig:LOChPT_xq_xdq}, where
the (2+1)-flavor and previous RBC quenched~\cite{Orginos:2005uy} and 2-flavor~\cite{Lin:2008uz} results are plotted together with the leading
nonlinear behavior predicted in heavy baryon chiral perturbation theory 
(HBChPT)~\cite{Chen:2001et, Arndt:2001ye,Detmold:2002nf}.  
\begin{eqnarray}
\langle x \rangle_{u-d}
&=&C\left [
1-\frac{3g^2_{A}+1}{(4\pi F_{\pi})^2}
m_{\pi}^2\ln\left(
\frac{m^2_{\pi}}{\mu^2}
\right)\right]\nonumber\\&&
+e(\mu^2)\frac{m_{\pi}^2}{(4\pi F_{\pi})^2}\\
\langle x \rangle_{\Delta u-\Delta d}
&=&\tilde{C}\left [
1-\frac{2g^2_{A}+1}{(4\pi F_{\pi})^2}
m_{\pi}^2\ln\left(
\frac{m^2_{\pi}}{\mu^2}
\right)\right]\nonumber\\&&
+\tilde{e}(\mu^2)\frac{m_{\pi}^2}{(4\pi F_{\pi})^2}
\end{eqnarray}
Although our lightest point may be beyond the applicability of HBChPT~\footnote{ 
It can be observed in some particular cases like the nucleon axial charge $g_A$ and the nucleon 
root-mean-squared (rms) charge radius~\cite{Lin:2008uz, Yamazaki:2009zq}.}, the downward trends are expected to develop at least in the vicinity of the physical pion mass point.
As a simple prediction, the curve is shown using experimental values for the nucleon axial charge and pion decay constant, $g_A=1.269$ and $F_\pi=92.8$ MeV, a chiral scale $\mu=m_N=940$ MeV, and by setting the unknown low energy constants to zero, $e(\mu)=\tilde e(\mu)=0$. The values in the chiral limit are obtained by requiring the curves to agree with experiment at the physical point.
It is interesting to note that the quenched results show no hint of this behavior, while the 2-flavor ones are inconclusive. We remind the reader that for the 2-flavor data, a smaller time separation between sources, $t_{\rm sep}\approx 1.16$ fm, was used. When the separation was increased for the lightest mass, the momentum and helicity fractions drop, but with a significant increase in the statistical error.

\begin{figure*}
\begin{center}
\includegraphics[width=\columnwidth,clip]{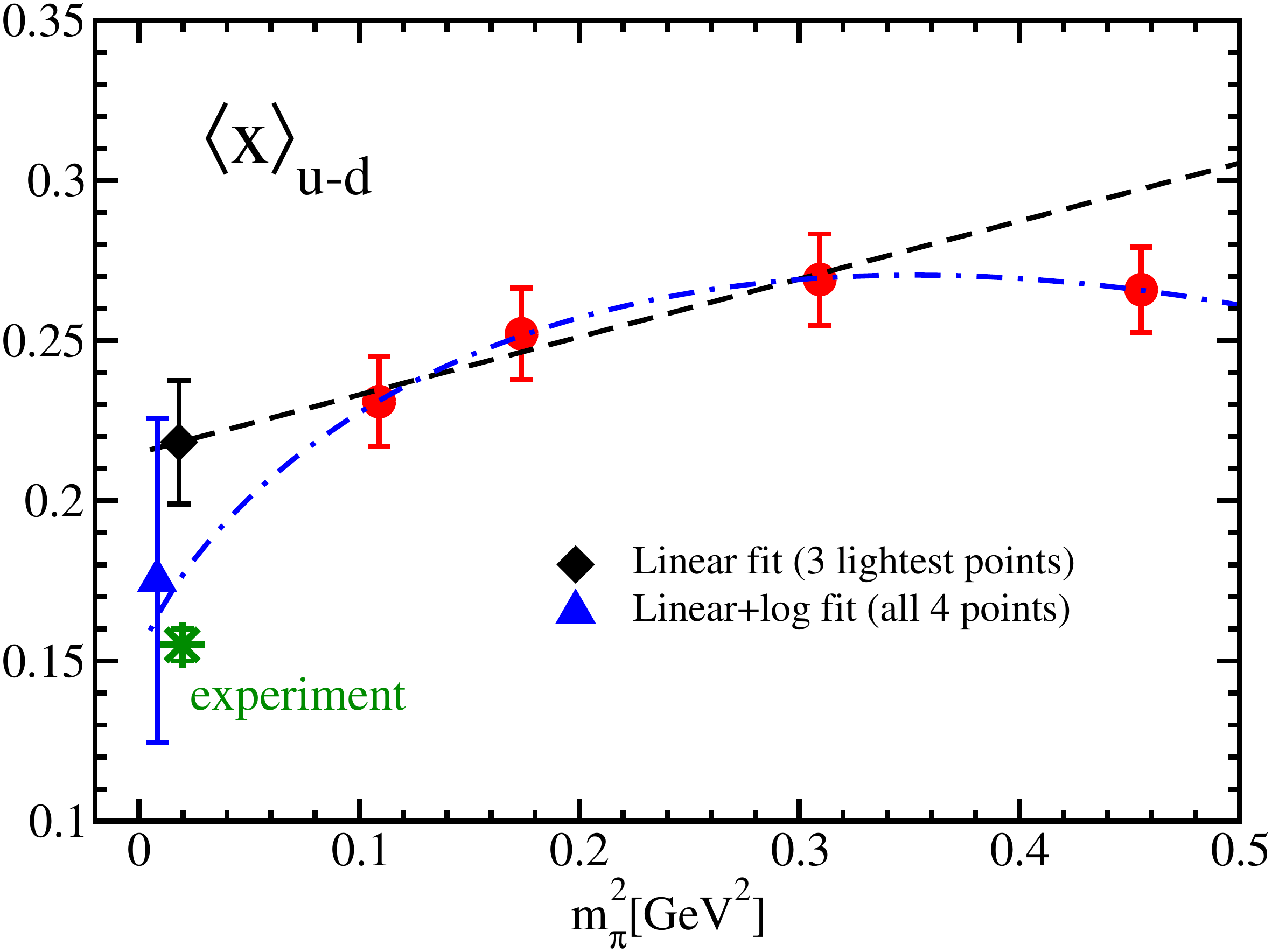}
\includegraphics[width=\columnwidth,clip]{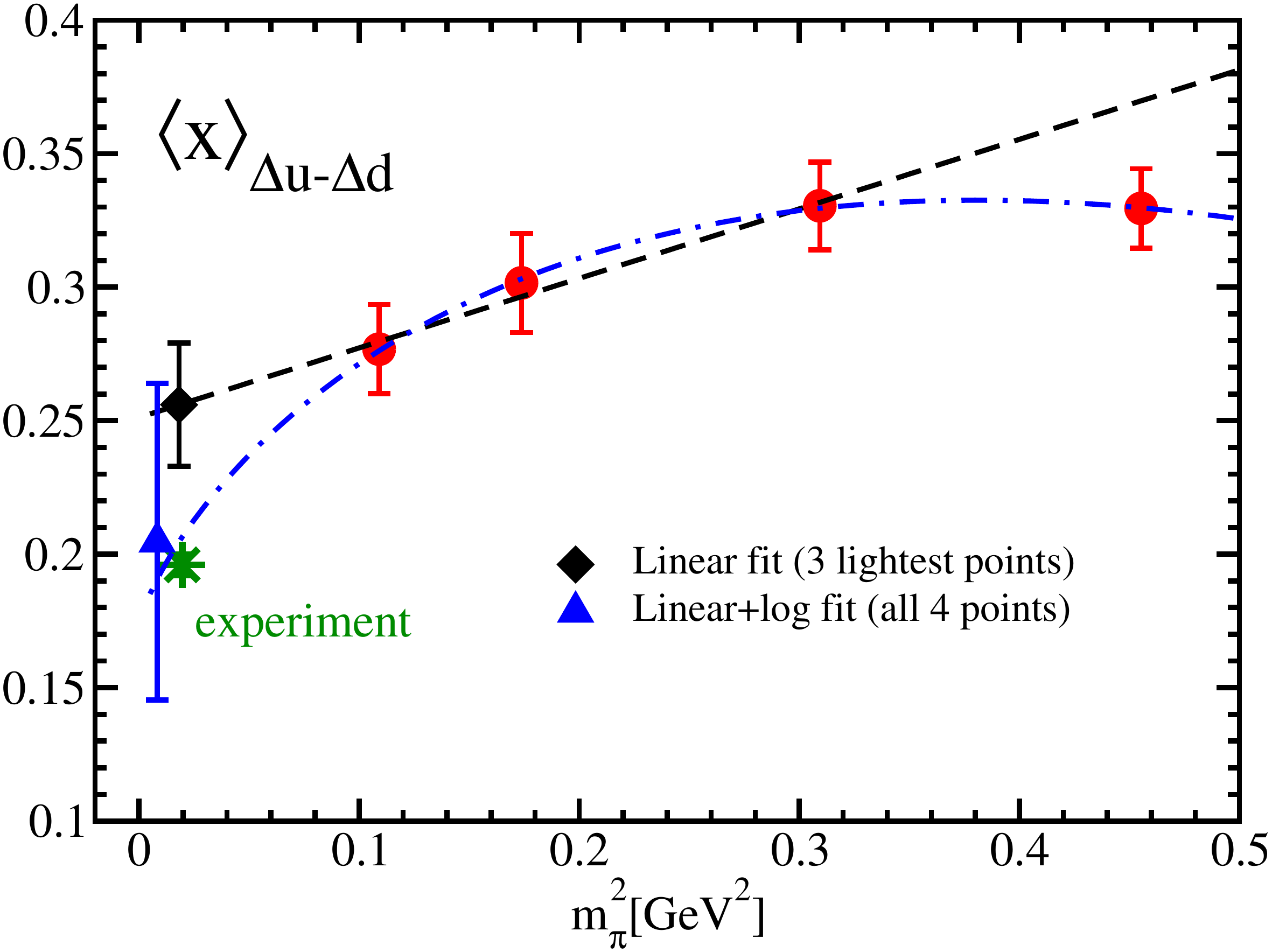}
\end{center}
\caption{Linear and leading HBChPT fits to the non-perturbatively renormalized quark momentum fraction, \( \langle x\rangle_{u-d}\),
and helicity fraction, \( \langle x \rangle_{\Delta u -\Delta d}\). The three lightest data points are 
used in the former, and all four in the latter. The diamond and triangle denote extrapolated values
at the physical point.}
\label{fig:cext_xq_xdq}
\end{figure*}

The physical point values  \(\langle x\rangle_{u-d}=0.218(19)\) and 
\(\langle x\rangle_{\Delta u-\Delta d}=0.256(23)\), determined
by simple linear chiral extrapolation of the three lightest points, overshoot the 
experimental values by more than 2-3 standard deviations as shown in Fig.~\ref{fig:cext_xq_xdq}.
On the other hand, nonlinear fit forms motivated by HBChPT,
\begin{equation}
\langle x \rangle_{q, \Delta q} (m_{\pi}^2)= c_0 + c_1 \cdot m_{\pi}^2 + d_1 m_{\pi}^2 \log(m_{\pi}^2),
\end{equation}
can easily accommodate all four data points and produce extrapolations 
to the physical point, in agreement with the respective experimental values, albeit with large statistical errors. 
Although this indicates a favorable trend, definitive results require simulations with 
several lighter quark masses than our lightest one.
All fits are summarized in Table~\ref{tab:Xud_DXud_phys}.

\subsection{Transversity (Tensor charge)}
\label{sec:tensor}

\begin{figure}
\begin{center}
\includegraphics[width=1.2\columnwidth,clip]{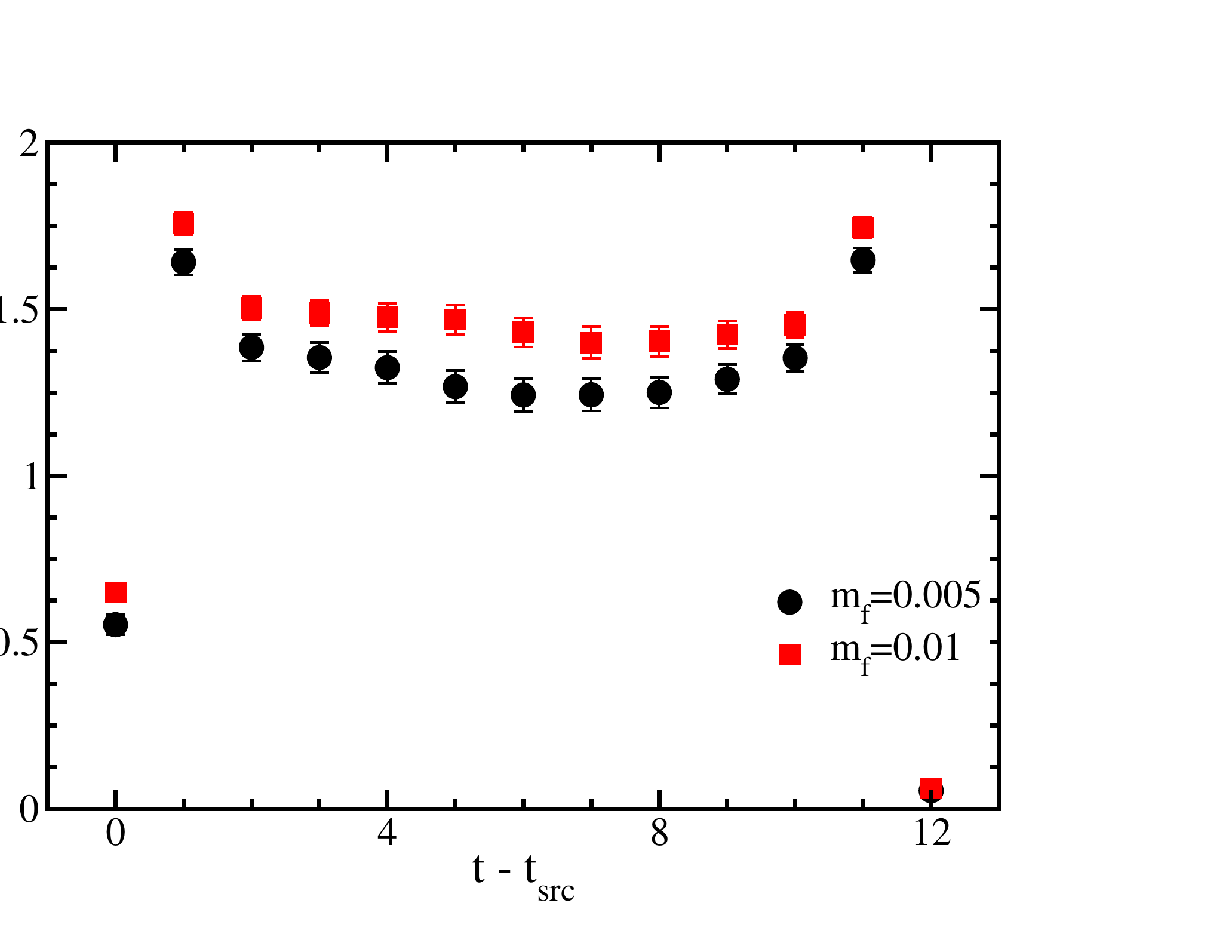}
\end{center}
\vskip -.5cm
\caption{Signals for the ratio of three- and two-point functions for the bare quark transversity, \(\langle 1 \rangle_{\delta u-\delta d}\). Quark mass 0.005 (circles) and 0.01 (squares).
\label{fig:1qsignal}}
\end{figure}

\begin{figure}
\begin{center}
\includegraphics[width=\columnwidth,clip]{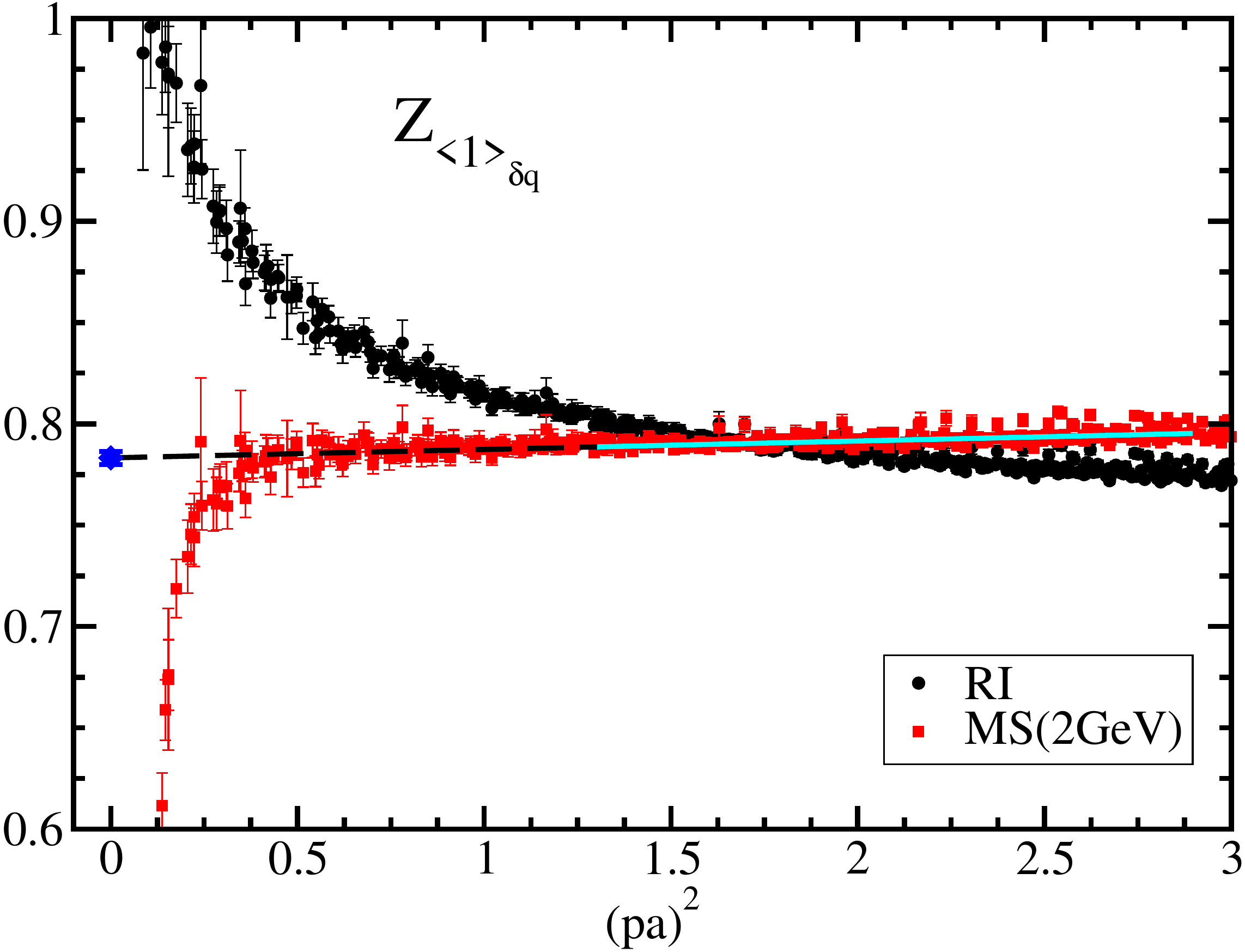}
\end{center}
\caption{Non-perturbative renormalization for the quark transversity, \(\langle 1 \rangle_{\delta u-\delta d}\). Circles denote the RI-MOM values, squares the $\overline{\rm MS}$ ones. The line denotes a linear fit used to remove the leading $O((ap)^2)$ lattice artifacts.\label{fig:1qnpr}}
\end{figure}

Results for the bare isovector tensor charge, \(\langle 1 \rangle_{\delta u - \delta d}\), are presented in Fig.\ \ref{fig:1qsignal}, and in Fig.\ \ref{fig:1qnpr} we present its non-perturbative renormalization constant. We obtain a renormalization factor of  \(Z^{\overline{\rm MS}}({\rm 2 GeV}) = 0.783(3)\).

\begin{figure}
\begin{center}
\includegraphics[width=1.1\columnwidth]{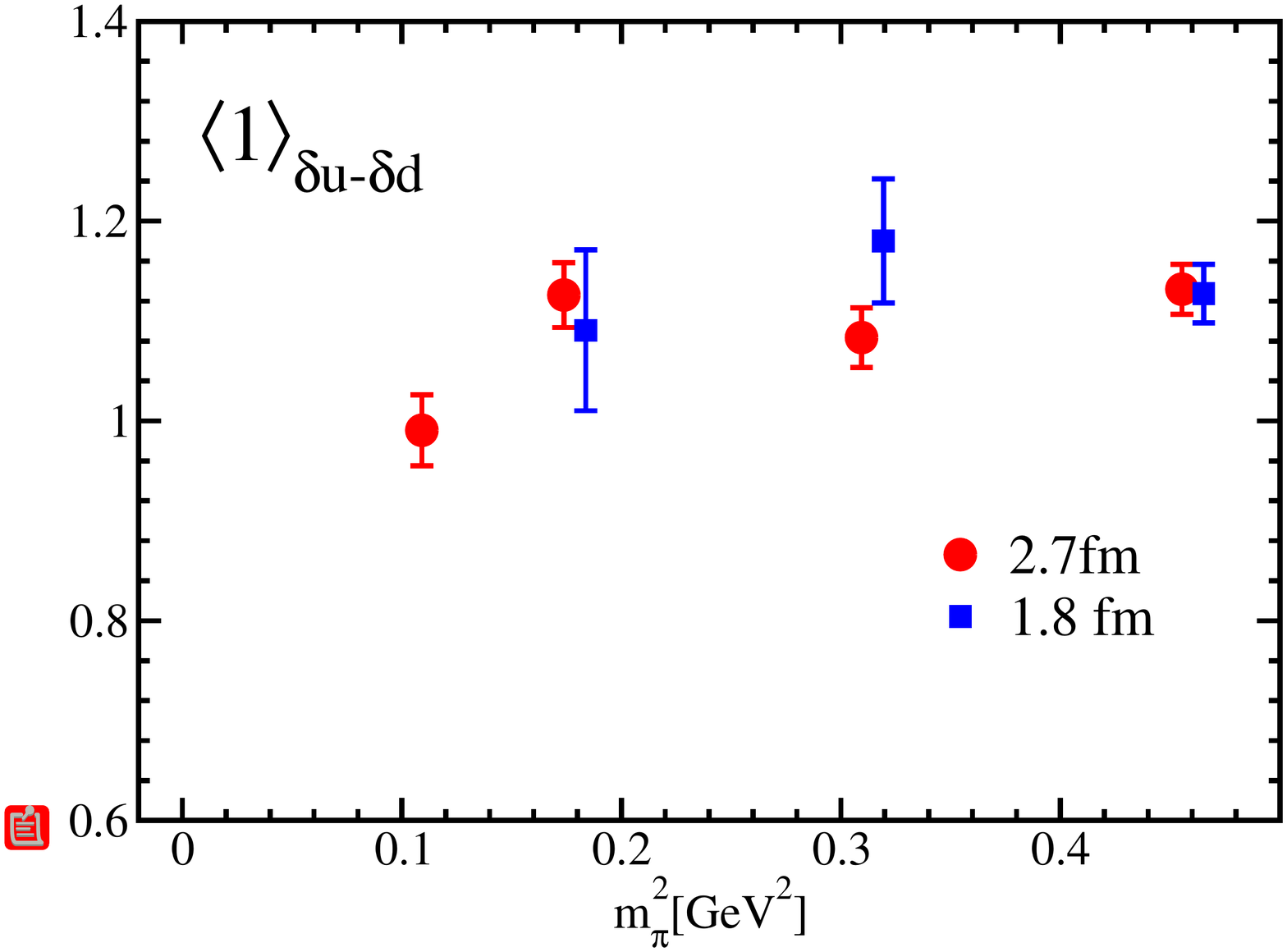}
\end{center}
\caption{Renormalizaed tensor charge, \(\langle 1 \rangle_{\delta u-\delta d}\). Both volumes are shown,  \(({\rm 2.7\  fm})^3\) (circles) and \(({\rm 1.8\ fm})^3\) (squares).
The square symbols have been moved slightly in the plus x-direction for clarity.~\label{fig:1q}}
\end{figure}

Combining them we obtain the renormalized tensor charge as presented in Fig.\ \ref{fig:1q} and summarized in Table~\ref{tab:All_L24}.
These provide a rough physical prediction which is still worthwhile since the experiments are yet to report a value.
If we fit the heavy three points with a constant we obtain a value of about 1.10(7).
Alternatively if we linearly extrapolate the two lightest points we would obtain about 0.7.

\subsection{Twist-3 moment}
\label{sec:d1}

\begin{figure}
\begin{center}
\includegraphics[width=\columnwidth,clip]{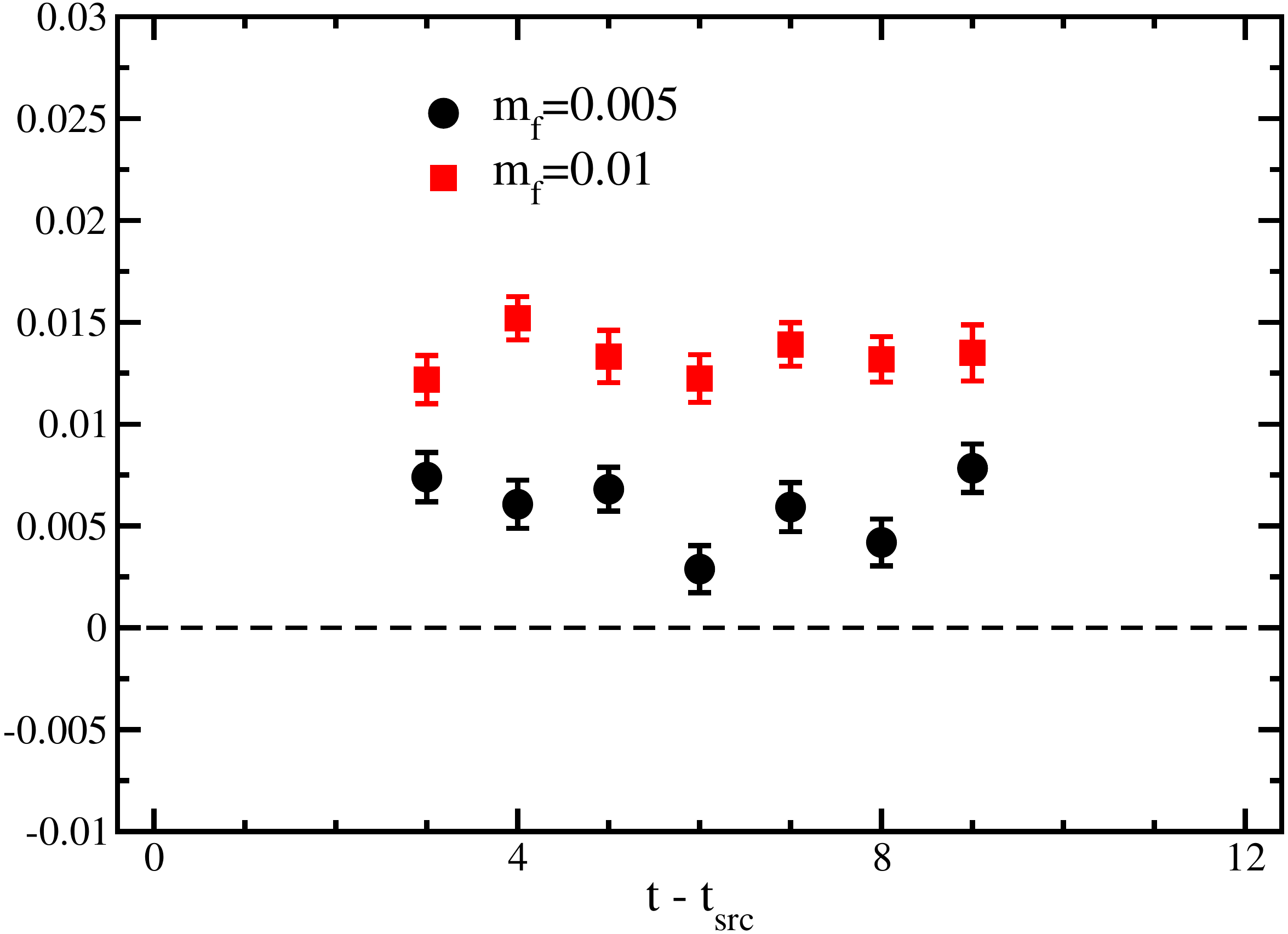}
\end{center}
\caption{
Bare signals from the ratio of three- and two-point function for twist-3 moment, $d_1$, for
$m_f=0.005$ and 0.01.~\label{fig:d1signal}}
\end{figure}

Figure \ref{fig:d1signal} presents the bare lattice signals for the twist-3 moment, \(d_1\), of the polarized structure function \(g_2\). They are summarized in Table~\ref{tab:twst3_L24}.

\begin{figure}
\begin{center}
\includegraphics[width=\columnwidth,clip]{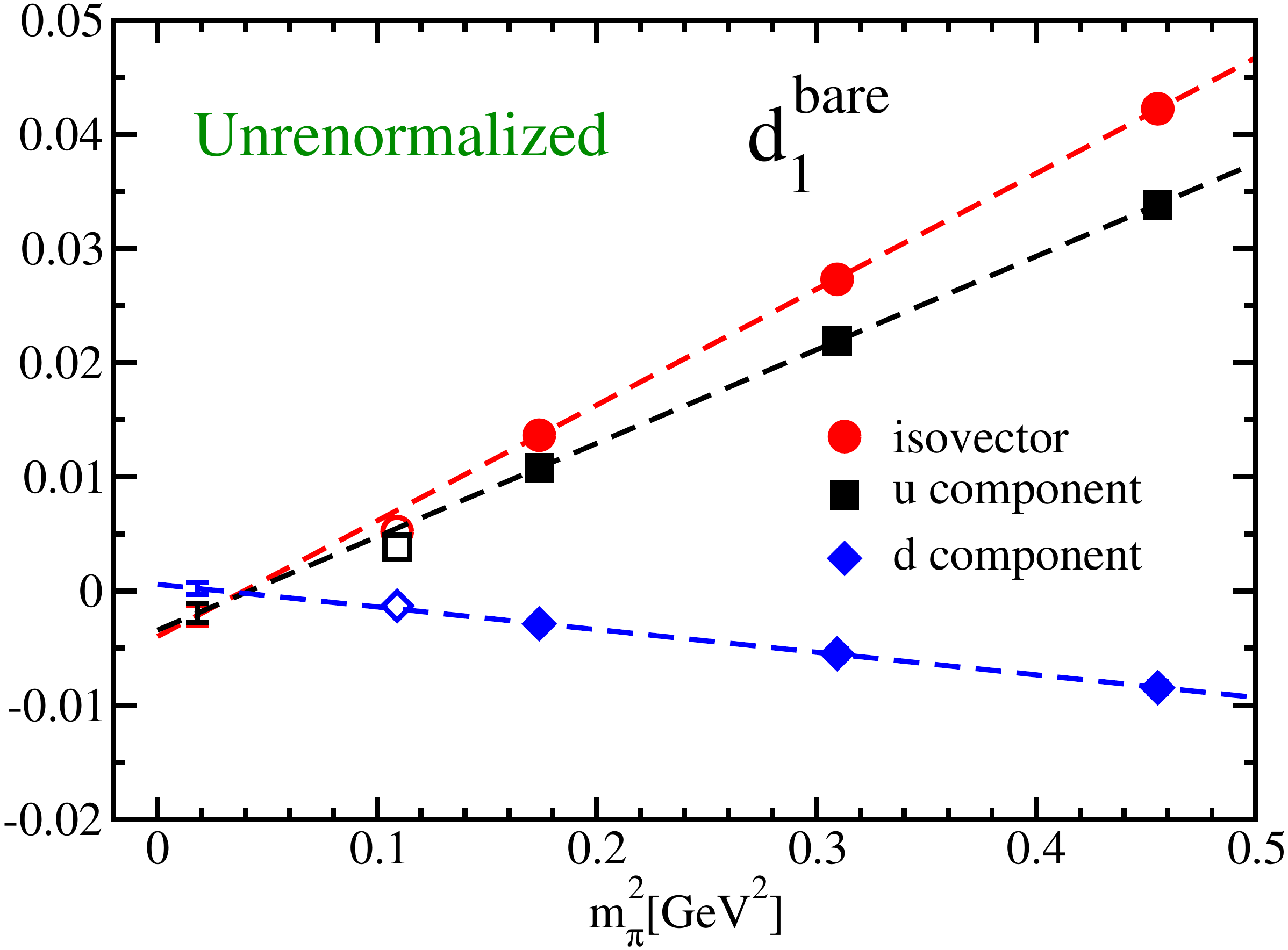}
\end{center}
\caption{Bare values of the twist-3 moment \( d_1 \), with
linear extrapolation to the physical point, excluding the lightest point (open symbols).
Up quark contribution (squares), down quark contribution (diamonds), and the isovector combination (circles) are shown. $24^3$ ensemble.
\label{fig:d1}}
\end{figure}

We have not yet  computed the renormalization constant for this quantity.
The quark-mass dependence of the bare values are presented in Fig.\ \ref{fig:d1}.
Our interest here is in whether the perturbatively obtained Wandzura-Wilczek relation~\cite{Wandzura:1977qf} holds.
From the smallness of the values obtained, we conclude it does.
We note that our results indicate that the lightest mass points deviate slightly from the linear trends set by the heavier points.

\begin{figure}
\begin{center}
\includegraphics[width=\columnwidth,clip]{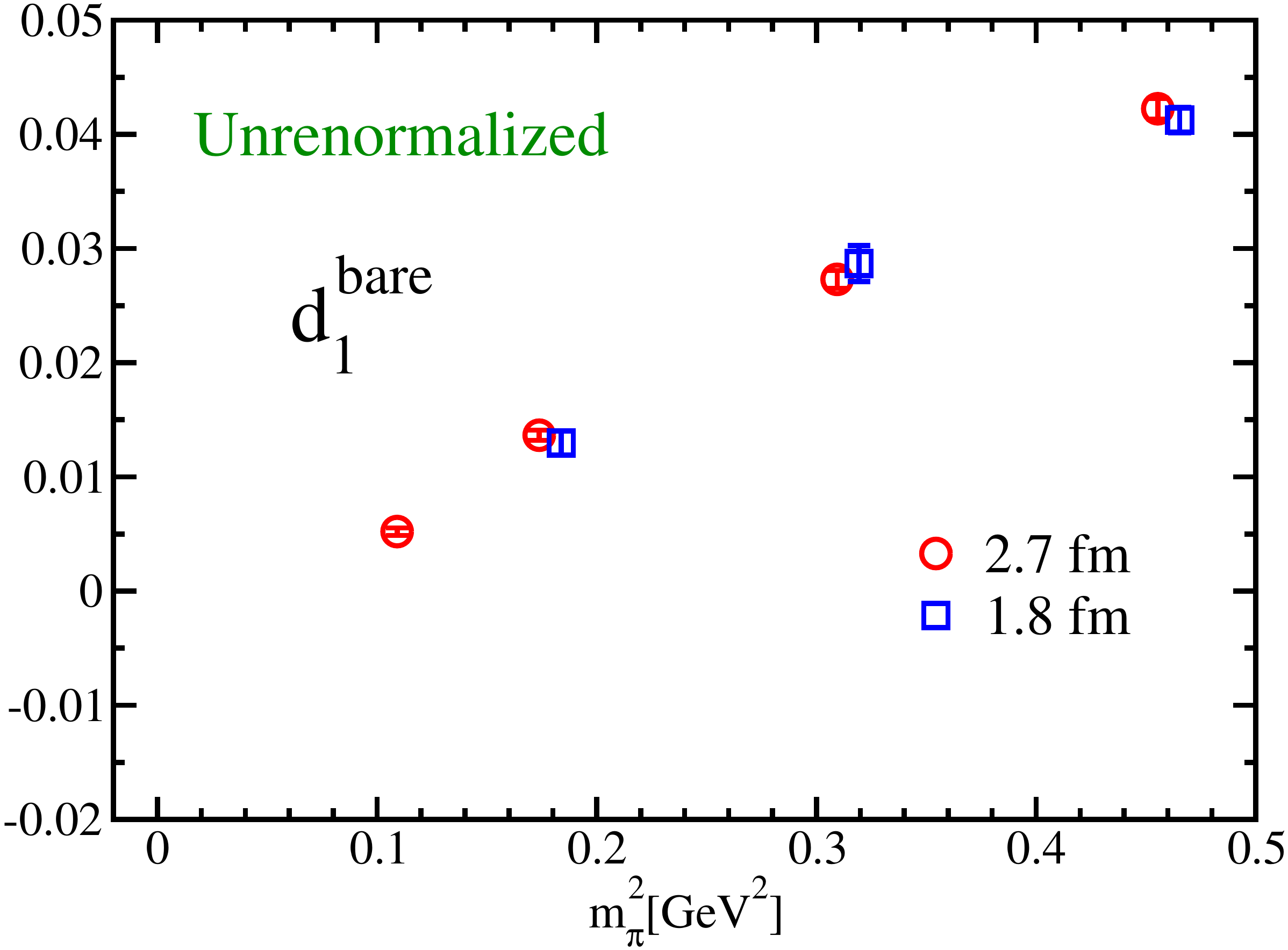}
\end{center}
\caption{
The twist-3 lowest moment of the polarized structure function \( d_1\) (not renormalized). Both volumes are shown,  \(({\rm 2.7\  fm})^3\) (circles) and \(({\rm 1.8\ fm})^3\) (squares). The square symbols have been moved slightly in the plus x-direction for clarity.\label{fig:d1_L}}
\end{figure}

In Fig.~\ref{fig:d1_L} values on different volumes are compared. $d_{1}$ appears to be insensitive to finite volume effects, at least in this range of light quark masses.

\section{Conclusions}
\label{sec:conclusions}

We have presented calculations of some of the lowest moments of nucleon structure functions in (2+1)-flavor QCD, using domain wall fermions and the Iwasaki gauge action. The calculations were carried out on two volumes at a single lattice spacing ($a^{-1}=1.73$ GeV) with quark masses that yield pion masses in the range 0.33 to 0.67 GeV. The results are encouraging.

The ratio of the bare quark momentum and helicity fractions, which is automatically renormalized, is found to be independent of the light quark mass through the range of our calculations, \(\frac{1}{5} m_{\rm strange} \le m_{\rm ud} \le \frac{3}{4} m_{\rm strange}\), and agrees with the value obtained from experiment within statistical error.
This is in contrast to a similarly automatically renormalized ratio, \(g_A/g_V\), of the isovector axialvector and vector charges that is severely distorted by the finite size of the lattice \cite{Yamazaki:2008py,Yamazaki:2009zq} at the lightest quark mass.

This suggests the corresponding downward trend toward the experimental values, as the quark mass decreases, of both the momentum, \(\langle x \rangle_{u-d}\), and helicity, \(\langle x \rangle_{\Delta u - \Delta d}\), fractions is a real physical effect.
Comparison of these results on two different volumes supports this observation. In fact, all of the moments studied here agree well, within statistical errors, on two different volumes, \((2.7\rm ~fm)^3\) and \((1.8\rm ~fm)^3\).

In addition to the momentum and helicity fractions, the non-perturbatively renormalized tensor charge, \(\langle 1 \rangle_{\delta u - \delta d}\), has been computed.
The chiral extrapolation, in particular, needs to be understood before an accurate prediction can be made. Since upcoming experiments have yet to report a value, we give a
rough estimate, from two different chiral extrapolations, that its value lies in the range 0.7-1.1.

\color{black}
The twist-3 moment of the \(g_2\) structure function, \(d_1\),  is also obtained.
Though yet to be renormalized, its smallness suggests the Wandzura-Wilczek relation holds.

The possibility that the long sought curvature of the moments in the chiral regime is becoming visible in our results has encouraged us to start calculations at even smaller quark masses ($m_{\pi}\approx 250$ and 180 MeV), on an even larger lattice ($L\approx 4.5$ fm). This ensemble, which is being generated by the RBC and UKQCD collaborations~\cite{Jung:2010jt}, was conceived, in part, 
to attain these goals for nucleon matrix elements.

\section*{Acknowlegements}
We thank the members of the RBC and UKQCD Collaborations. HL is supported by the US DOE under grant DE-FG03-97ER4014, JZ by STFC grant ST/F009658/1,
SO thanks the RIKEN-BNL Research Center for partial support,
SS is supported by the JSPS for a Grant-in-Aid for Scientific Research (C), No. 19540265,
TB by the US DOE under contract DE-FG02-92ER40716, and
YA by the JSPS for a Grant-in-Aid for Scientific Research (C), No. 21540289.
RIKEN, BNL, the U.S. DOE, Edinburgh University and the UK PPARC provided facilities essential for this work. The computations reported here were carried out on the QCDOC supercomputers at the RBRC and the University of Edinburgh.

\clearpage

\begin{table}[!t]
\caption{Summary of renormalization factors in the \( \overline{\rm MS} \) scheme
at 2 GeV in the chiral limit.
}
\begin{center}
\begin{tabular}{llll} \hline\hline
\multicolumn{1}{c}{\( m_f \)} & 
\multicolumn{1}{c}{\( {\cal O}_{44}^q \)} & 
\multicolumn{1}{c}{ \( {\cal O}_{\{34\}}^{5q} \)}& 
\multicolumn{1}{c}{\( {\cal O}_{34}^{\sigma q} \)} \\
\hline
\( -m_{\rm res} \) & 1.149(45) & 1.154(35) &0.783(3)\\
\hline\hline
\end{tabular}
\end{center}
\label{table:renorm}
\end{table}

\begin{table}[!t]
\caption{
Bare quark momentum and helicity fractions and their naturally renormalized ratio 
on the \( (2.7\ {\rm fm})^3\) ensemble.}
\label{tab:Xud_L24}
\begin{center}
\begin{tabular}{lllc}
\hline\hline
$m_f$  &
\multicolumn{1}{c}{$\langle x \rangle_{u-d}$} &
\multicolumn{1}{c}{$\langle x \rangle_{\Delta u-\Delta d}$}&
\multicolumn{1}{c}{$\langle x \rangle_{u-d}/\langle x \rangle_{\Delta u-\Delta d}$}\\
\hline
0.005  & 0.201(9) & 0.240(13) & 0.835(46) \\
0.01    & 0.219(9) & 0.261(14) & 0.842(42) \\
0.02    & 0.234(8) & 0.286(11) & 0.821(40) \\
0.03    & 0.231(7) & 0.285(10) & 0.807(32) \\
\hline\hline
\end{tabular}
\end{center}
\end{table}

\begin{table}[!t]
\caption{
Bare quark momentum and helicity fractions and their naturally renormalized ratio 
on the \( (1.8\ {\rm fm})^3\) ensemble.}
\label{tab:Xud_L16}
\begin{center}
\begin{tabular}{lllc}
\hline\hline
$m_f$  &
\multicolumn{1}{c}{$\langle x \rangle_{u-d}$} &
\multicolumn{1}{c}{$\langle x \rangle_{\Delta u-\Delta d}$}&
\multicolumn{1}{c}{$\langle x \rangle_{u-d}/\langle x \rangle_{\Delta u-\Delta d}$}\\
\hline
0.01 & 0.221(18) & 0.263(29) & 0.808(89)\\
0.02 & 0.256(14) & 0.291(22) & 0.875(56)\\
0.03 & 0.236(7) & 0.300(11) & 0.784(27)\\
\hline\hline
\end{tabular}
\end{center}
\end{table}

\begin{table}[!t]
\caption{
Isovector combination ($u-d$) of the quark momentum fraction \( \langle x \rangle_{q}\), helicity fraction
\( \langle x \rangle_{\Delta q}\) 
and transversity \( \langle 1 \rangle_{\delta q}\),
non-perturbatively renormalized in the \(\overline{\rm MS}\) scheme at 2 GeV.}
\label{tab:All_L24}
\begin{center}
\begin{tabular}{lccc}
\hline\hline
$m_f$  &
\multicolumn{1}{c}{$\langle x \rangle_{u-d}^{\overline{\rm MS}}(2 {\rm GeV})$} &
\multicolumn{1}{c}{$\langle x \rangle_{\Delta u- \Delta d}^{\overline{\rm MS}}(2 {\rm GeV})$} &
\multicolumn{1}{c}{$\langle 1 \rangle_{\delta u- \delta d}^{\overline{\rm MS}}(2 {\rm GeV})$} \\
\hline
0.005 & 0.231(14) & 0.277(17)
&1.265(45) \\
0.01 & 0.252(14) & 0.302(19)
&1.438(41) \\
0.02 & 0.269(14) & 0.330(16)
&1.384(38) \\
0.03 & 0.266(13) & 0.329(15)
&1.446(32) \\
\hline\hline
\end{tabular}
\end{center}
\end{table}

\begin{table}[!t]
\caption{
Summary of extrapolations to the physical point of the renormalized first moment of the quark momentum
and helicity fractions. 
\label{tab:Xud_DXud_phys}}
\begin{center}
\begin{tabular}{lcc}
\hline\hline
& $\langle x \rangle_{u-d}^{\overline{\rm MS}}(2{\rm GeV})$ & $\langle x \rangle_{\Delta u -\Delta d}^{\overline{\rm MS}}(2{\rm GeV})$ \\
\hline
Linear vs $m_{\pi}^2$ (3 points) & 0.218(19) & 0.256(23) \\
ChPT vs $m_{\pi}^2$ (4 points) & 0.175(51) & 0.205(59) \\
\hline
Experiment & 0.154(3) & 0.196(4) \\
\hline\hline
\end{tabular}
\end{center}
\end{table}

\begin{table}[!t]
\caption{
Bare twist-3 first moment of the polarized structure function, $d_1$, on the \( (2.7\ {\rm fm})^3\) ensemble.}
\label{tab:twst3_L24}
\begin{center}
\begin{tabular}{lccc}
\hline\hline
$m_f$  &
\multicolumn{1}{c}{$d_1^{u-d}$} &
\multicolumn{1}{c}{$d_1^{u}$} &
\multicolumn{1}{c}{$d_1^{d}$} \\
\hline
0.005 & 0.0052(3) & 0.0038(3)
&$-0.0013(2)$ \\
0.01 & 0.0137(4) & 0.0108(4)
&$-0.0029(3)$ \\
0.02 & 0.0273(8) & 0.0219(8)
&$-0.0055(4)$ \\
0.03 & 0.0422(9) & 0.0338(9)
&$-0.0085(5)$ \\
\hline\hline
\end{tabular}
\end{center}
\end{table}

\bibliography{nuc_ref}

\end{document}